\documentclass[a4paper,10pt,twocolumn]{article}
\usepackage{listings}
\usepackage{embedfile}
\usepackage{graphicx}
\usepackage{tikz}
\usepackage{pgfplots}
\usepackage{hyperref}
\usepackage{caption}
\usepackage{subcaption}
\usepackage{marvosym}
\usepackage[margin=2cm]{geometry}
\usepackage[utf8]{inputenc}
\usepackage{lmodern}
\usepackage{color}
\usepackage{longtable}
\usepackage{array}
\usepackage{makeidx}
\usepackage{amsfonts}
\usepackage{amsmath}
\usepackage{abstract}
\newcommand{\pdf}{\textsc{pdf}}
\newcommand{\pyt}{\emph{Python}}
\newcommand{\hmi}{\textsc{hmi}}
\newcommand{\wsg}{\textsc{wysiwyg}}
\newcommand{\mde}{\textsc{mde}}
\renewcommand{\u}{$\sqcup$}
\newcommand{\prereq}{Default}\def\wordsl#1{\wordsloopiter#1 \nil} \def\wordsloopiter#1 #2\nil{ \textcolor{brown}{\langle #1 \rangle} \ifx&#2& \let\next\relax \else \def\next{\wordsloopiter#2\nil} \fi \next}
\newcommand{\req}[3]{\paragraph{{\sc Req.} {\tiny \textcolor{blue}{$\langle \prereq{}#1 \rangle$} } } {\em #2 $\circ$ } \hfill {\tiny $\wordsl{#3}$ } \index{Req[\prereq{}#1]}}
\title{\bf \'{E}conomie des biens immatériels \\ {\large le réseau $\sqcup${\underline{$net$}}}}
\author{Laurent Fournier {\small \{\url{laurent.fournier@cupfoundation.net}\}}  \\ 
{\tiny \tt [\texttt{tDavy}]} }

\makeindex
\begin{document}
\lstset{language=Python, breaklines=true}
\captionsetup{labelfont=sc}
\hypersetup{colorlinks,linkcolor=Red, urlcolor=Blue}
\usetikzlibrary{shapes,fit,arrows,shadows,backgrounds,svg.path}
\newcommand\net{$\sqcup${\underline{$net$}}}
\twocolumn[
\maketitle
\begin{onecolabstract}
We introduce a new economic system suited for {\em Intangible Goods} ({\sc ig}). We argue that such system can now be implemented in the real world using advance technics in distributed network computing and cryptography. The specification of the so called \net{} is presented. To Limit the number of financial transactions, the system is forced to define its own currency, with many benefits. The new ``{\em cup}'' currency, extended worldwide, is dedicated to {\sc ig}, available only for person-to-person trading, protected from speculation and adapted for tax recovery with no additional computation.
Those nices features makes the \net{} a new democratic tool, fixing specific issues in {\sc ig} trading and reviving a whole domain activity.
We emphasis on the fact that all proposed documentation, algorithm, program in any language related to this proposal shall be open-source without any possibility to post any patent of any sort on the system or subsystem. This new trading model should be considered as a pure intellectual construction, like parts of Mathematics and then belongs to nobody or everybody, like $1+1=2$. Next step will be to test, validate the security of various implementations details, and to ask for legal rules adaptations. \\
The first draft paper is written in French language and posted to {\bf arXiv.org} and {\bf hal.archive-ouverte.fr}. We expect to provide an English translation before Christmas.

\vspace{.4cm}\par\indent {\small {\bf Keywords\/}: economics, intangible good, anti-rival good, peer-to-peer, strong-authentication, download-freeness, digital brain, intellectual property.}
\end{onecolabstract}
]
\section{Introduction}
L'\'{E}conomie porte principalement son attention sur les biens {\em matériels} et biens {\em rivaux}, faisant de la notion de {\it rareté} un de ses fondements. De tels biens ont la propriété suivante~: produire un deuxième exemplaire n'est pas à coût nul et demande un effort en ressources consommées. Des économies d'échelle font que produire un million d'exemplaires du même produit ne revient pas à un million de fois le coût de production d'un seul exemplaire, mais ce coût est inévitablement supérieur à celui d'un seul exemplaire. Nous définissons comme {\em biens immatériels} ({\em Intangible Goods} - {\sc ig}) des produits, objets de commerce\footnote{les e-mails, réseaux sociaux et autres blogs, bien qu'immatériels, ne constituent pas une valeur marchande.}, mais n'ayant pas la propriété précédente. Ainsi, les {\sc ig}s coûtent exactement le même prix que ce soit pour en produire un seul, deux, dix ou un million. On dit aussi que leur {\em coût marginal} est nul. La notion de rareté ne s'applique pas sur les {\sc ig}s et dès lors qu'ils sont créés, il peuvent exister indéfiniment. Ces biens sont qualifiés de {\em biens non rivaux} et même de {\em bien anti-rivaux} de par le caractère moutonnier des consommateurs.
On ne s'interresse pas ici à d'autres entités immatérielles comme les {\em services}, qui ont un effet toujours limité dans le temps, ni aux {\em actifs immatériels} dont l'objet est de mesurer la richesse d'une entreprise à des fins de rachat, emmission d'action ou de protection de la propriété intellectuelle. Les {\sc ig}s sont ici des biens marchand, visibles, désirables par tout un chacun, grand public ou initié.
Intuitivement, les {\sc ig}s sont des \oe{}uvres culturelles, des articles, des essais, des romans, des journaux, de la musique, des photos, des films, des reportages, des jeux vidéo,\ldots tout objet fruit d'un travail, d'une performance intellectuelle et matérialisé sur un support mémoire électronique. On néglige ici le coût effectif du composant mémoire. Ce coût est très faible, il est réparti et facilement supporté par les usagers créateurs ou les utilisateurs. Par exemple, un roman de cinq cent pages\cite{bellanger} pouvant être l'\oe{}uvre de plusieurs années de travail ne représente qu'un mégaoctet de données, soit moins d'un dixième de centime d'euros en coût mémoire.

Il est indubitable que dans notre société post-industrielle, le nombre d'{\sc ig}s augmente sensiblement\footnote{principalement poussé par la consommation de masse et la démocratisation de bien culturels.} et que le réseau {\em Internet} participe largement à leur diffusion.

Cependant, le commerce des {\sc ig}s est encore basé sur le même modèle économique et commercial\footnote{ce modèle n'a pas changé depuis l'invention de la monnaie, vers 680 av.{\sc j.-c.} et seule l'introduction de la notion d'intérêt comme rémunération d'un prêt fixe un contrat pour une durée finie entre un vendeur et un acheteur.} que celui des biens matériels, à savoir qu'à un instant donné un individu -- le vendeur -- propose pour son compte ou celui d'une personne morale un bien à un niveau de prix qui puisse intéresser des acheteurs potentiels, mais la transaction s'effectue toujours entre un seul vendeur et un seul acheteur et ne dure qu'un très court instant. Le prix est quasiment indépendant du nombre d'acheteurs si bien qu'un {\sc ig} populaire peut rapidement faire la fortune du vendeur, de l'auteur et des intermédiaires dans la chaine de distribution.

Au premier abord, on ne trouve rien à redire à ce système. Il souffre pourtant d'une inadaptation flagrante à la spécificité des {\sc ig}s.

Si quelqu'un achète un {\sc ig}, par exemple sur {\em Internet}, il doit légitimement exiger que ce bien soit consommable n'importe o\`{u} et n'importe quand. Quelque soit le PC, SmartPhone, tablette, {\sc tv},....qu'il utilise, que ce matériel soit ou non sa propriété, il devrait pouvoir bénéficier, lire, voir, écouter, consommer cet {\sc ig}. Cette exigence est loin d'être remplie, même si elle commence à être satisfaite dans des systèmes très fermés (produits {\em Apple Inc.} et {\em iCloud} en particulier).
En conséquence, la copie de fichiers contenant ces {\sc ig}s ne peut pas être totalement interdite, ne serait ce que pour les transferts d'un appareil à l'autre. Notons que l'action de copie est contraignante pour l'utilisateur qui désire avant tout bénéficier de l'{\sc ig}. Comme il n'est pas plus difficile de copier un {\sc ig} entre appareils d'une même personne ou entre appareils appartenant à des personnes différentes, la diffusion et le partage, légal ou illégal des {\sc ig}s est un phénomène largement observé, malgré les lois.

Du reste, comme les auteurs et autres intermédiaires proposent des prix basés sur le modèle économique classique des supports matériels (papier, photo, cd, dvd), ou qu'ils désirent compenser les effets de la copie illégale, l'offre légale reste très chère, incitant dans un cercle vicieux l'usager à se tourner vers la copie illégale, ce qui divise encore plus la population.

Dans ce modèle classique, un auteur d'{\sc ig} qui remporte un certain succès peut gagner (ou faire gagner aux intermédiaires) un revenu quasi infini sans jamais en rétrocéder une partie aux acheteurs, raison pour laquelle certain refusent en bloc le système et se procurent des {\sc ig}s par copie illégale.  

Il n'y a pas non plus aujourd'hui de système vraiment sécurisé qui assure à un consommateur ayant perdu ou effacé un {\sc ig} de le retrouver facilement sans le racheter.

D'une façon générale, pour établir un échange juste, et donc plus bénéfique\footnote{sans toutefois proposer ici un critère objectif d'optimisation.} pour un nombre maximum d'individus, il serait nécessaire que le revenu tiré d'un bien soit proportionnel au travail et au génie de son concepteur. Il y a visiblement désaccord entre une minorité qui pensent que le génie peut être infini, pouvant justifier un revenu sans limite, et une grande majorité de citoyens, voulant bien valoriser le travail intellectuel et artistique, dans les limites raisonnables de la finitude des êtres humains\footnote{Un humoriste dirait qu'il est bien difficile de conduire deux {\em Ferrari} en même temps!}.

Imaginons maintenant que l'on puisse définir un système économique plus juste par la seule concession sur le revenu potentiellement infini des auteurs, il est fort probable que le marché des {\sc ig}s augmenterait en volume, mais aussi que globalement les (véritables) auteurs seraient en moyenne mieux et plus justement rémunérés. Enfin, la tentation de contourner le système déclinerait pour mettre tout le monde (ceux qui achetaient à prix fort et ceux qui copiaient gratuitement) sur un pied d'égalité devant le même bien culturel, représenté par le même {\sc ig}.
Nous verrons aussi que dans ce système, la copie est inutile, que les effacements accidentels peuvent se corriger, sans racheter le bien, en sus d'autres avantages qui permettraient d'annihiler la fraude.
\begin{figure}
\centering
\tikzset{
  b/.style={rectangle,rounded corners=4pt,draw=none,text=black,fill=blue!15},
  g/.style={circle,draw=none,text=white,fill=red!40},
  mya/.style={->, very thick,line width=3pt,draw=gray}, 
  myv/.style={->, very thick,line width=3pt,draw=black!40!green},
  myr/.style={->, thick,draw=gray, dashed},
}
\subcaptionbox{Tangible Good, one-to-one relation}{
\begin{tikzpicture}[align=center]
    \node[b] (a) at (0,0) {Buyer\\{\huge \Smiley}};
    \node[g] (mg) at (2,0) {\bf \sc tg};
    \node[b] (v) at (4,0) {Seller\\{\huge \Smiley}};
    \draw[myv] (v) -- (mg); \draw[mya] (a) -- (mg);
\end{tikzpicture}
}
\subcaptionbox{Intangible Good, one-to-many relation}{
\begin{tikzpicture}[align=center]
    \node[b] (v) at (1,0) {{\huge\Smiley\Mobilefone}\\Seller};
    \node[g] (ig) at (2,2) {\bf \sc ig};
    \node[b] (a1) at (0,4) {$1^{st}$ buyer\\{\huge\Smiley\Mobilefone}}; 
    \node[b] (a2) at (2,4) {$2^{nd}$ buyer\\{\huge\Smiley\Mobilefone}};
    \node[b] (a3) at (4,4) {$3^{rd}$ buyer\\{\huge\Smiley\Mobilefone}};
    \node[b] (a4) at (6,3) {\ldots buyer\\{\huge\Smiley\Mobilefone}};
    \draw[myv] (v) -- (ig);
    \draw[mya] (a1) -- (ig);\draw[mya] (a2) -- (ig);\draw[mya] (a3) -- (ig);\draw[mya] (a4) -- (ig);
    \draw[myr] (a2) -- (a1);\draw[myr] (a3) -- (a2);\draw[myr] (a4) -- (a3);
    \draw[myr] (a1) -- (v);\draw[myr] (a2) -- (v);\draw[myr] (a3) -- (v);\draw[myr] (a4) -- (v);
\end{tikzpicture}
}
\caption{\it {\rm ({\sc a})} Le modèle économique classique adapté aux biens matériels et né l'an 680 av.{\sc j.-c.}, sans besoin de réseau numérique. {\rm ({\sc b})} Le nouveau modèle économique adapté aux biens immatériels et basé sur le réseau numérique \net{}, les liens en pointillés représentent les flux automatiques en cup. Le smartphone est le principal matériel d'authentification et de consommation des {\sc ig}s.}
\end{figure}
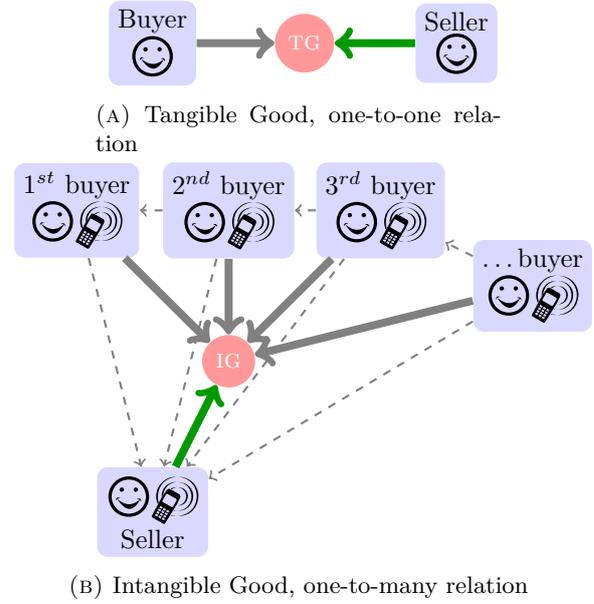
\section{Le modèle}
Notre proposition de nouveau modèle économique pour les bien immatériels ({\sc ig}) est basé sur le principe des prix et revenu bornés et de transparence du nombre d'acheteurs.
Ainsi un auteur ne propose pas un prix pour un {\sc ig} à une seule valeur, mais à deux valeurs, ou un couple de valeurs $(\mathcal{P}_1, \mathcal{I}_\infty)$, noté $\textcolor{blue}{\boxed{{{\bf \mathcal{P}_1}}\sqcup^{\mathcal{I}_\infty}}}$\footnote{le symbole $\sqcup$ "{\em square cup}" ou "{\em cup}" sert aussi de séparateur des deux nombres lors de la prononciation d'un prix à l'oral.}.\\
La première valeur $\mathcal{P}_1$ indique le prix du tout premier acheteur; sachant qu'une grande partie lui sera remboursé par la suite.
La deuxième valeur $\mathcal{I}_\infty$\footnote{on a toujours $\mathcal{I}_\infty \gg \mathcal{P}_1$} est le revenu ({\em $\mathcal{I}$ncome}) total reçu par le créateur du bien s'il vend à un nombre très grand d'acheteurs. Cette dernière valeur peut être élevée, mais l'auteur renonce à un revenu quasi infini. L'argent économisé par l'ajout au modèle d'un revenu maximal est automatiquement reversé aux précédents acheteurs. La transaction n'est plus instantanée entre deux personnes, elle est initiée à un instant mais n'a pas de fin dans le temps (Figures \ref{fig_both} et \ref{fig_price}) et surtout, elle est établie entre un vendeur et un ensemble croissant d'acheteurs.
Concrètement, le vendeur va voir son compte en banque crédité de $\mathcal{P}_1$ initialement puis lentement progresser jusqu'à tendre vers $\mathcal{I}_\infty$. Le premier acheteur payera $\mathcal{P}_1$ et verra ensuite aussi son compte en banque crédité de petite sommes jusqu'à tendre vers un remboursement presque total de son achat initial. Le processus est le même pour les autres acheteurs au détail près que le prix initial $\mathcal{P}_n$ est fonction du nombre $n$ d'acheteurs le précédant ($\mathcal{P}_n < \mathcal{P}_1$). Pour les derniers acheteurs, $\mathcal{P}_n$ est très proche et même égal à zéro, après arrondi.

On comprends intuitivement que plus il y a d'acheteurs et plus on tend vers la limite donnant satisfaction à chacun, en toute transparence; l'auteur est rémunéré au maximum de ce qu'il prévoyait et les acheteurs disposent du bien pour un prix borné et décroissant jusqu'à presque rien. De plus, à tout moment, tous les acheteurs d'un même {\sc ig} payent exactement le même prix.
Enfin, comme le vendeur fait connaître explicitement son revenu maximal escompté, l'acheteur potentiel peut évaluer la popularité de l'{\sc ig} et donc estimer le montant du remboursement, ainsi que sa vitesse. 

Mettons ce principe en équation.

Les prix (acheteurs) et revenu {\em Income} (vendeur) sont respectivement définis par les séries mathématique $\mathcal{P}$ et $\mathcal{I}$ fonction toutes les deux du nombre $n$ d'acheteurs du même {\sc ig}.
On a toujours~:
$$\mathcal{I}_n = n\mathcal{P}_n \hspace{2em} \forall n \in \mathbb{N}^*$$ 

Le revenu $\mathcal{I}$ est une suite croissante ayant les propriétés suivantes~:
$$\mathcal{I}_1=\mathcal{P}_1 \hspace{3em} \lim_{n\to+\infty} \mathcal{I}_n = \mathcal{I}_\infty$$
 
Le prix $\mathcal{P}$ est une suite décroissante ayant les propriétés suivantes~:
$$\mathcal{P}_1=\mathcal{I}_1  \hspace{3em}  \lim_{n\to+\infty} \mathcal{P}_n = 0$$ 

\subsection{Le modèle linéaire par morceaux}
\begin{figure}
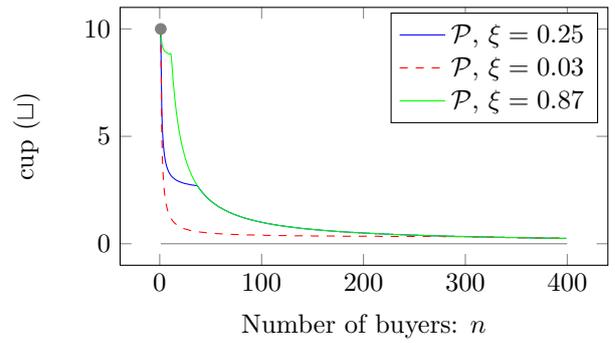

%
\caption{\it La courbe de prix de la transaction $\textcolor{blue}{\boxed{{\bf 10}\sqcup^{\mbox{\tiny 100}}}}$. L'inflexion marque la fin de revenu pour le vendeur, et tout nouvel achat est redistribué aux précédents acheteurs jusqu'à tendre vers un prix nul. La suite reste décroissante quelque soit la valeur de $\xi \in [0,1]$}
\label{fig_price}
\end{figure}

Une solution possible pour $\mathcal{I}$ et $\mathcal{P}$ est définie par récursion~:

\begin{equation} \mathcal{I}_n = \mathcal{I}_{n-1} + \xi\mathcal{P}_1\end{equation}

si $\mathcal{I}_n < \mathcal{I}_\infty$ sinon $\mathcal{I}_n=\mathcal{I}_\infty$, avec $\xi$ un coefficient d'équilibre entre la progression du revenu et la redistribution entre acheteurs.
$\xi$ est un paramètre système dont la valeur est dans l'intervalle $[0, 1]$ (Figure \ref{fig_price}).\\
\begin{itemize}
\item Si $\xi = 0$, le revenu final est donc égal au prix initial, les acheteurs se répartissent le coût et le vendeur ne reçoit plus de revenu dès le deuxième achat. La transaction est alors similaire à la condition $\mathcal{I}_\infty = \mathcal{P}_1$.\\
\item Si $\xi = 1$, alors $\mathcal{P}_n = \mathcal{P}_1 \hspace{1em} \forall n, \mathcal{I}_n < \mathcal{I}_\infty$. Le vendeur est rémunéré le plus rapidement possible et les acheteurs ne sont remboursés seulement quand le vendeur a atteint le maximum de revenu.
\end{itemize}
Le revenu et le prix peuvent s'exprimer alors simplement en fonction du nombre d'acheteurs $n$.

\begin{equation} \mathcal{I}_n = \mathcal{P}_1 (1+ \xi(n-1)) \hspace{2em} \forall n \in \mathbb{N}^* , \mathcal{I}_n < \mathcal{I}_\infty \end{equation}
\begin{equation} \mathcal{P}_n = \mathcal{P}_1 \left(\xi +  \frac{ \left( 1-\xi\right) }{n} \right)  \end{equation}

Quand $\mathcal{I}_n = \mathcal{I}_\infty$, on a $\mathcal{P}_n = \mathcal{I}_\infty/n$.

Comme désiré, le vendeur reçoit un revenu effectif dans l'intervalle $[\mathcal{P}_1,\mathcal{I}_\infty]$ et les acheteurs payent un prix dans l'intervalle $[0,\mathcal{P}_1]$, et même $[0,\mathcal{P}_n]$ avec $\mathcal{P}_n<\mathcal{P}_1$ (Figure \ref{fig_both}). 
Pour l'achat numéro $n$, la somme à rembourser $\mathcal{R}_n$ au précédents acheteurs, ainsi que le surplus de revenu $\Delta_n$ touché par le vendeur sont~:
\begin{equation} 
\Delta_n = \xi\mathcal{P}_1 \hspace{3em} 
\mathcal{R}_n = \frac{\mathcal{P}_n - \Delta_n }{n-1} = \frac{(1-\xi)\mathcal{P}_1}{n(n-1)}
\end{equation}
Entre deux opérations financières séparées de $k+1$ achats, toujours au bout de l'achat numéro $n$, la valeur de gain de revenu et celle de remboursement sont~:
\begin{equation} \Delta_{n,k} = (k+1)\xi\mathcal{P}_1 \end{equation}
\begin{equation} \mathcal{R}_{n,k} = \sum_{i=n-k}^n{\mathcal{R}_{i}} = (1-\xi)\mathcal{P}_1 \sum_{i=n-k}^n \frac{1}{i(i-1)}\end{equation}

avec toujours~:
$$\mathcal{I}_n = n\mathcal{P}_n \hspace{2em} \forall n \in \mathbb{N}^*$$ 
Un utilisateur désirant affirmer son mécénat envers un {\sc ig} particulier ou envers un auteur particulier peut décider d'acheter non pas un mais plusieurs ($k+1$) exemplaires au même instant.
Le prix qu'il devra acquitter sera $\forall n \in \mathbb{N}^* , \mathcal{I}_n < \mathcal{I}_\infty$:
\begin{equation} \mathcal{P}_{n,k} = \mathcal{P}_1 \left( \xi(k+1) + (1-\xi) \sum_{i=n-k}^n \frac{1}{i} \right) \end{equation}
sinon~:
\begin{equation} 
\mathcal{P}_{n,k} = \mathcal{I}_\infty \sum_{i=n-k}^n \frac{1}{i}
\end{equation}
avec~:
$$ \mathcal{P}_{n,k} < k \mathcal{P}_{n-k+1} < k \mathcal{P}_{n-k}$$

\subsection{Le modèle exponentiel}
Une solution possible pour $\mathcal{I}$ est définie par~:
\begin{equation} \mathcal{I}_n = \mathcal{I}_\infty +(\mathcal{P}_1-\mathcal{I}_\infty) e^{-k\xi(n-1)}\end{equation}
avec
\begin{equation} k = \ln(\mathcal{I}_\infty - \mathcal{P}_1) -\ln(\mathcal{I}_\infty - 2\mathcal{P}_1) \end{equation}
Le modèle exponentiel a les mêmes propriétés que le modèle linéaire, le revenu $\mathcal{I}$ est une suite croissante vers $\mathcal{I}_\infty$ et le prix $\mathcal{P}$ est une suite décroissante vers $0$.\\
Les figures \ref{diff_model} et \ref{diff_modelxi1} montrent la différence entre les deux modèles pour la même valeur de $\xi$.

\subsection{Enoncé du principe}
Les deux modèles précédents respectent un principe général applicable à tout bien immatériels. Ce principe s'énonce litéralement par~:

\emph{Le revenu cumulé d'un vendeur d'un bien immatériel est borné, tous les acheteurs ont payés à tout instant le même prix cumulé, décroissant vers zéro.}

Mathématiquement~:
$$

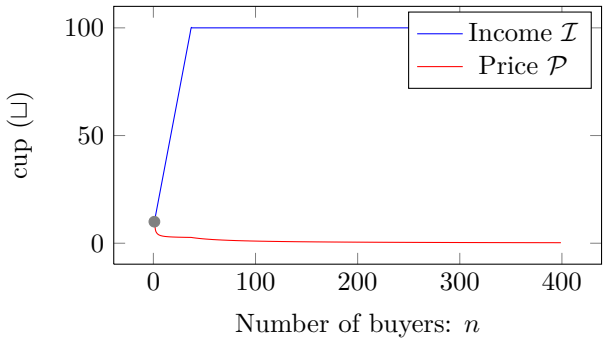
\caption{\it La transaction $\textcolor{blue}{\boxed{{\bf 10}\sqcup^{\mbox{\tiny 100}}}}$ représentée par les courbes de revenu $\mathcal{I}$ et de prix $\mathcal{P}$ sur le même graphique}
\label{fig_both}
\end{figure}

\begin{figure}
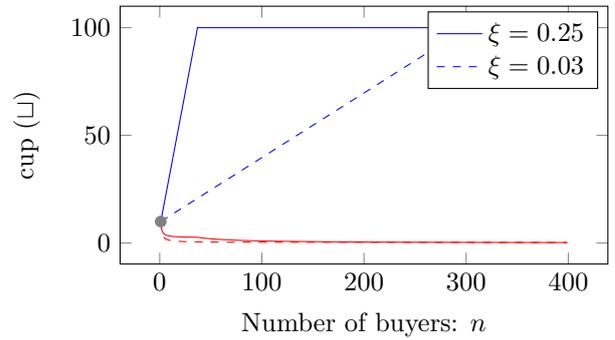

%
\caption{\it Influence du paramètre $\xi$ sur les courbes de revenu et de prix, pour la transaction $\textcolor{blue}{\boxed{{\bf 10}\sqcup^{\mbox{\tiny 100}}}}$}
\label{fig_xi}
\end{figure}

\begin{figure}
%
\caption{\it Influence du revenu maximum sur les courbes de revenu et de prix, pour les transactions $\textcolor{blue}{\boxed{{\bf 10}\sqcup^{\mbox{\tiny 100}}}}$ et $\textcolor{blue}{\boxed{{\bf 10}\sqcup^{\mbox{\tiny 60}}}}$ pour la même valeur de $\xi$}
\label{fig_max_income}
\end{figure}

\begin{figure}
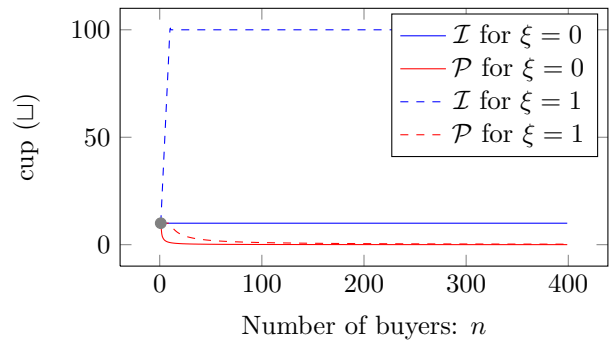

%
\caption{\it Les courbes de prix et revenu de la transaction $\textcolor{blue}{\boxed{{\bf 10}\sqcup^{\mbox{\tiny 100}}}}$ pour les valeurs extrêmes de $\xi$}
\label{fig_extreme}
\end{figure}

\subsection{La monnaie}
Nous avons donc remplacé la transaction une-à-une ({\em 1-1}), convenable pour les biens matériels, par une transaction une-vers-plusieurs ({\em 1-n}), mieux adaptée aux {\sc ig}s (Figure \ref{fig_both}).
 
A première vue, notre système requiert beaucoup plus d'opérations financières que le système classique si chaque achat d'un {\sc ig} est suivi de $n$ opérations entre le vendeur et chacun des acheteurs, afin d'une part de rajouter un bénéfice au revenu du vendeur, et d'autre part de rembourser les précédents acheteurs, sans oublier de débiter le compte du tout dernier acheteur. Comme l'acte d'achat d'un {\sc ig} n'est jamais fini dans le temps, on pourrait aussi craindre une saturation du système à cause des multiples opérations financières élémentaires.

La complexité des transactions est en $\mathcal{O}(n)$ pour le système classique et en $\mathcal{O}(n^2)$ pour notre proposition.

Nous répondons à cette problématique par la création d'une nouvelle monnaie dédiée exclusivement aux {\sc ig}s. Appelons "{\em square cup}" ou "{\em cup}", notée $\sqcup$ le nom et l'unité de cette monnaie. L'idée est que tous les échanges entre vendeurs et acheteurs d'{\sc ig}s se font désormais en {\em cup} avec certaines contraintes mais de nombreux avantages;

La notation complète d'un prix dans cette monnaie est par exemple~:
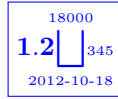
$$\textcolor{blue}{\boxed{{\bf 1.2}\!\!\!\!\!\!\!\bigsqcup_{\mbox{\tiny 2012-10-18}}^{\mbox{\tiny 18000}}\!\!\!\!\!\!\!\mbox{\tiny 345}}}$$
indique que l'{\sc ig} à été mis à la vente le 18 Octobre 2012, sa date de naissance au format {\sc iso}, qu'il y a déjà eu 345 acheteurs, que le prix courant est de $1.2$ {\em cup} et que le vendeur réclame un revenu final de $18000$ {\em cup}\footnote{pour faciliter l'internationalisation, le mot {\em cup} ne s'accorde pas en nombre en Français comme dans les autres langues.}.
On pourra noter un prix plus simplement, $\textcolor{blue}{\boxed{{\bf 1.2}\sqcup}}$ ou $\textcolor{blue}{\boxed{{\bf 1.2}\sqcup^{\mbox{\tiny18000}}}}$ seulement si le matériel de visualisation permet d'afficher les attributs manquant quand l'{\sc ig} est pré-sélectioné.

Si un éditeur désire vendre un support mémoire ou une impression papier contenant un {\sc ig}. Ce {\sc cd}, {\sc dvd}, ou livre est un bien matériel pouvant contenir l'impression de la référence de l'{\sc ig} avec le prix indiqué pour le premier acheteur. En effet, le nombre d'acheteurs à un instant donné ne peut être connu à l'impression du support et donc on donnera le prix avant le tout premier achat correspondant à un nombre d'acheteurs égal à zéro. Dans ce cas, on peut ne pas faire apparaître le $0$ dans l'indication du prix.
$$\textcolor{blue}{\boxed{{\bf 1.2}\!\!\!\!\!\!\!\bigsqcup_{\mbox{\tiny 2012-10-18}}^{\mbox{\tiny 18000}}\!\!\!\!\!\!\!\mbox{\tiny 345}}} \equiv
\textcolor{blue}{\boxed{{\bf 15.7}\!\!\!\!\!\!\!\bigsqcup_{\mbox{\tiny 2012-10-18}}^{\mbox{\tiny 18000}}\!\!\mbox{\tiny 0}}} \equiv
\textcolor{blue}{\boxed{{\bf 15.7}\!\!\!\!\!\!\!\bigsqcup_{\mbox{\tiny 2012-10-18}}^{\mbox{\tiny 18000}}}} $$

Voici les règles imposées par cette monnaie~:
\begin{itemize}
\item Un individu ne peut pas posséder un compte débiteur en {\em cup}; tout achat est conditionné au provisionnement suffisant du compte de l'acheteur.
C'est un peu similaire à une carte de téléphone prépayée, donnant droit à un nombre de minutes fixé de communication.
Chaque individu est responsable de transférer régulièrement une somme en {\em cup} depuis un compte dans la monnaie locale de son pays.
\item les comptes en {\em cup} n'autorisent pas les crédits, et l'épargne n'est pas rémunérée.
\item les {\em cup} ne servent exclusivement qu'à acheter des {\sc ig}s, directement à leur auteurs. Il ne peut y avoir d'intermédiaire.
\end{itemize}

Voyons maintenant les avantages de recourir à cette nouvelle monnaie~:
\begin{itemize}
\item Comme les acteurs de notre système sont des individus du monde entier, les transactions sont facilitées sans opérations de change et donc sans risque de spéculation sur les cours des monnaies entre \'{E}tats.

\item Le lien entre les comptes en {\em cup} et les comptes bancaires réels des intervenants est initié sur demande de ces derniers.
Chaque acheteur demande à intervalle régulier de provisionner son compte en {\em cup} et de débiter son compte bancaire dans la monnaie locale et chaque vendeur demande quand il le désire le transfert de ses {\em cup} dans une monnaie locale. L'opération se réalise par la création d'une page de document au format {\sc pdf/a} constituant une sorte d'{\sc ig} à faire valoir à sa banque et ayant la fonction de chèque bancaire ou de titre de créance.

\item Les opérateurs chargés de ces transferts sont des organismes bancaires validés par un \'{E}tat ou union d'\'{E}tat. Contrairement aux individus, les banques peuvent fournir des {\em cup} aux demandeurs sur le système, contre le débit d'un montant sur leur compte bancaire. Inversement, les banques peuvent acheter des {\em cup} à un auteur et le créditer d'une somme correspondante dans sa monnaie locale. 

\item Le taux de change d'achat et de vente n'est pas obligatoirement le même, ce pour mettre en place une taxe, sorte de {\sc tva}, facilement collectable par l'Administration. Ces taux sont fixés par les états ou responsables de la monnaie et non par les banques elles mêmes, qui ne doivent recevoir aucune commission. Ne sont pas soumis à cette taxe les achats culturels réalisés par des créateurs, car tout reste en {\em cup}. Un véritable marché des biens culturels peut s'intaller avec une contrainte très forte que la revente n'est pas possible par le protocole du réseau \net. Seul l'auteur original peut vendre son \oe{}uvre.

\item L'\'{E}tat peut aussi mettre en place des opérations particulières à destination d'une partie de la population, par exemple, fournir gratieusement un nombre de {\em cup} aux jeunes ou à certains foyers à revenu modeste afin de leur faciliter l'accès aux biens culturels. 

\end{itemize}
 
Comme toutes ces opérations en {\em cup} sont automatisées, sans aucune intervention humaine, les banques ne peuvent pas revendiquer la prise d'une commission, si minime soit-elle.

Il est important de noter que seul l'auteur d'un {\sc ig} peut le vendre, il n'est pas possible de revendre un {\sc ig}. Si par exemple, un individu change la signature numérique d'un {\sc ig} et s'en attribue la paternité, il encours des poursuites pénales s'il y a copie flagrante, et comme chaque acteur est fortement authentifié, le voleur ne peut se cacher derrière l'anonymat.
La revente interdite bloque aussi toute tentative de spéculation sur la monnaie en {\em cup}.
Rappelons qu'il y a pas d'intérêts adossables à un compte en {\em cup}. Le solde sur le compte ne sert qu'a lisser les achats et ventes courantes, mais il n'y a aucun intérêt à faire une épargne. 

Seul le créateur/auteur d'un {\sc ig} peut connaître la liste des acheteurs et la date de leur achat, pour ce même {\sc ig}. Seul un acheteur a accès à son historique d'achat. Ces listes sont chiffrées. Cependant, une personne peut vouloir communiquer à une autre une partie de sa liste d'achat afin que cette dernière commande les mêmes {\sc ig}s. C'est même le nouveau moyen de communiquer avec ce système. Tout partage de liste est sous la responsabilité du propriétaire, qui ne pourra pas se plaindre légalement de la divulgation de cette information. Un acheteur ou un vendeur peut aussi fournir à la justice une preuve, au sens mathématique ce ces listes, pour sa défense ou sur demande explicite d'un juge. Ces cas sont possibles si des individus tentent de vendre un {\sc ig} volé ou-bien ayant un contenu illicite.   

Dans la réalité, les calculs sur les prix et revenus ne se font pas avec une précision infinie. Si l'on admet d'arrondir au centime du {\em cup}, alors il se produit le phénomène que pour un {\sc ig} populaire, une partie de la population l'aura acheté un centime de {\em cup} tandis que l'autre (les derniers) l'acquièrent gratuitement. Cela concrétise une forme de passage dans le domaine public de la création. L'inégalité reste minime et reste compensée par l'avantage de bénéficier de l'{\sc ig} plus tôt, avant qu'il ne devienne public. 

Nous résumons la différence entre le modèle des biens matériels et celui des {\sc ig} par le tableau \ref{table_diff}.

Notre nouveau modèle économique n'a de sens que s'il est réalisable sur un ensemble de points fondamentaux. 
Notre thèse de ce document est justement d'expliquer que les avancées en informatique, réseaux et cryptographie permettant de réaliser ce système, un prototype dans un premier temps et de le déployer s'il donne satisfaction.
    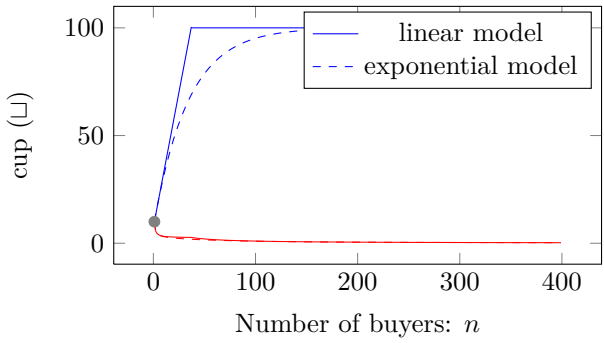
\begin{figure}
\begin{tikzpicture}
  \begin{axis}[ xlabel=Number of buyers: $n$, ylabel=cup ($\sqcup$), width=8cm, height=5cm]
    \addplot[smooth,blue] plot coordinates { (1,10.00) (2,12.50) (3,15.00) (4,17.50) (5,20.00) (6,22.50) (7,25.00) (8,27.50) (9,30.00) (10,32.50) (11,35.00) (12,37.50) (13,40.00) (14,42.50) (15,45.00) (16,47.50) (17,50.00) (18,52.50) (19,55.00) (20,57.50) (21,60.00) (22,62.50) (23,65.00) (24,67.50) (25,70.00) (26,72.50) (27,75.00) (28,77.50) (29,80.00) (30,82.50) (31,85.00) (32,87.50) (33,90.00) (34,92.50) (35,95.00) (36,97.50) (37,100.00) (38,100.00) (39,100.00) (40,100.00) (41,100.00) (42,100.00) (43,100.00) (44,100.00) (45,100.00) (46,100.00) (47,100.00) (48,100.00) (49,100.00) (50,100.00) (51,100.00) (52,100.00) (53,100.00) (54,100.00) (55,100.00) (56,100.00) (57,100.00) (58,100.00) (59,100.00) (60,100.00) (61,100.00) (62,100.00) (63,100.00) (64,100.00) (65,100.00) (66,100.00) (67,100.00) (68,100.00) (69,100.00) (70,100.00) (71,100.00) (72,100.00) (73,100.00) (74,100.00) (75,100.00) (76,100.00) (77,100.00) (78,100.00) (79,100.00) (80,100.00) (81,100.00) (82,100.00) (83,100.00) (84,100.00) (85,100.00) (86,100.00) (87,100.00) (88,100.00) (89,100.00) (90,100.00) (91,100.00) (92,100.00) (93,100.00) (94,100.00) (95,100.00) (96,100.00) (97,100.00) (98,100.00) (99,100.00) (100,100.00) (101,100.00) (102,100.00) (103,100.00) (104,100.00) (105,100.00) (106,100.00) (107,100.00) (108,100.00) (109,100.00) (110,100.00) (111,100.00) (112,100.00) (113,100.00) (114,100.00) (115,100.00) (116,100.00) (117,100.00) (118,100.00) (119,100.00) (120,100.00) (121,100.00) (122,100.00) (123,100.00) (124,100.00) (125,100.00) (126,100.00) (127,100.00) (128,100.00) (129,100.00) (130,100.00) (131,100.00) (132,100.00) (133,100.00) (134,100.00) (135,100.00) (136,100.00) (137,100.00) (138,100.00) (139,100.00) (140,100.00) (141,100.00) (142,100.00) (143,100.00) (144,100.00) (145,100.00) (146,100.00) (147,100.00) (148,100.00) (149,100.00) (150,100.00) (151,100.00) (152,100.00) (153,100.00) (154,100.00) (155,100.00) (156,100.00) (157,100.00) (158,100.00) (159,100.00) (160,100.00) (161,100.00) (162,100.00) (163,100.00) (164,100.00) (165,100.00) (166,100.00) (167,100.00) (168,100.00) (169,100.00) (170,100.00) (171,100.00) (172,100.00) (173,100.00) (174,100.00) (175,100.00) (176,100.00) (177,100.00) (178,100.00) (179,100.00) (180,100.00) (181,100.00) (182,100.00) (183,100.00) (184,100.00) (185,100.00) (186,100.00) (187,100.00) (188,100.00) (189,100.00) (190,100.00) (191,100.00) (192,100.00) (193,100.00) (194,100.00) (195,100.00) (196,100.00) (197,100.00) (198,100.00) (199,100.00) (200,100.00) (201,100.00) (202,100.00) (203,100.00) (204,100.00) (205,100.00) (206,100.00) (207,100.00) (208,100.00) (209,100.00) (210,100.00) (211,100.00) (212,100.00) (213,100.00) (214,100.00) (215,100.00) (216,100.00) (217,100.00) (218,100.00) (219,100.00) (220,100.00) (221,100.00) (222,100.00) (223,100.00) (224,100.00) (225,100.00) (226,100.00) (227,100.00) (228,100.00) (229,100.00) (230,100.00) (231,100.00) (232,100.00) (233,100.00) (234,100.00) (235,100.00) (236,100.00) (237,100.00) (238,100.00) (239,100.00) (240,100.00) (241,100.00) (242,100.00) (243,100.00) (244,100.00) (245,100.00) (246,100.00) (247,100.00) (248,100.00) (249,100.00) (250,100.00) (251,100.00) (252,100.00) (253,100.00) (254,100.00) (255,100.00) (256,100.00) (257,100.00) (258,100.00) (259,100.00) (260,100.00) (261,100.00) (262,100.00) (263,100.00) (264,100.00) (265,100.00) (266,100.00) (267,100.00) (268,100.00) (269,100.00) (270,100.00) (271,100.00) (272,100.00) (273,100.00) (274,100.00) (275,100.00) (276,100.00) (277,100.00) (278,100.00) (279,100.00) (280,100.00) (281,100.00) (282,100.00) (283,100.00) (284,100.00) (285,100.00) (286,100.00) (287,100.00) (288,100.00) (289,100.00) (290,100.00) (291,100.00) (292,100.00) (293,100.00) (294,100.00) (295,100.00) (296,100.00) (297,100.00) (298,100.00) (299,100.00) (300,100.00) (301,100.00) (302,100.00) (303,100.00) (304,100.00) (305,100.00) (306,100.00) (307,100.00) (308,100.00) (309,100.00) (310,100.00) (311,100.00) (312,100.00) (313,100.00) (314,100.00) (315,100.00) (316,100.00) (317,100.00) (318,100.00) (319,100.00) (320,100.00) (321,100.00) (322,100.00) (323,100.00) (324,100.00) (325,100.00) (326,100.00) (327,100.00) (328,100.00) (329,100.00) (330,100.00) (331,100.00) (332,100.00) (333,100.00) (334,100.00) (335,100.00) (336,100.00) (337,100.00) (338,100.00) (339,100.00) (340,100.00) (341,100.00) (342,100.00) (343,100.00) (344,100.00) (345,100.00) (346,100.00) (347,100.00) (348,100.00) (349,100.00) (350,100.00) (351,100.00) (352,100.00) (353,100.00) (354,100.00) (355,100.00) (356,100.00) (357,100.00) (358,100.00) (359,100.00) (360,100.00) (361,100.00) (362,100.00) (363,100.00) (364,100.00) (365,100.00) (366,100.00) (367,100.00) (368,100.00) (369,100.00) (370,100.00) (371,100.00) (372,100.00) (373,100.00) (374,100.00) (375,100.00) (376,100.00) (377,100.00) (378,100.00) (379,100.00) (380,100.00) (381,100.00) (382,100.00) (383,100.00) (384,100.00) (385,100.00) (386,100.00) (387,100.00) (388,100.00) (389,100.00) (390,100.00) (391,100.00) (392,100.00) (393,100.00) (394,100.00) (395,100.00) (396,100.00) (397,100.00) (398,100.00) (399,100.00)  };\addlegendentry{linear model}
    \addplot[smooth,blue, dashed] plot coordinates { (1,10.00) (2,12.61) (3,15.15) (4,17.61) (5,20.00) (6,22.32) (7,24.58) (8,26.76) (9,28.89) (10,30.95) (11,32.96) (12,34.90) (13,36.79) (14,38.62) (15,40.41) (16,42.13) (17,43.81) (18,45.44) (19,47.03) (20,48.56) (21,50.06) (22,51.51) (23,52.91) (24,54.28) (25,55.61) (26,56.89) (27,58.14) (28,59.36) (29,60.54) (30,61.68) (31,62.80) (32,63.87) (33,64.92) (34,65.94) (35,66.93) (36,67.89) (37,68.82) (38,69.73) (39,70.60) (40,71.46) (41,72.28) (42,73.09) (43,73.87) (44,74.63) (45,75.36) (46,76.08) (47,76.77) (48,77.45) (49,78.10) (50,78.74) (51,79.35) (52,79.95) (53,80.53) (54,81.10) (55,81.65) (56,82.18) (57,82.70) (58,83.20) (59,83.69) (60,84.16) (61,84.62) (62,85.07) (63,85.50) (64,85.92) (65,86.33) (66,86.73) (67,87.11) (68,87.48) (69,87.85) (70,88.20) (71,88.54) (72,88.88) (73,89.20) (74,89.51) (75,89.82) (76,90.11) (77,90.40) (78,90.68) (79,90.95) (80,91.21) (81,91.47) (82,91.71) (83,91.95) (84,92.19) (85,92.41) (86,92.63) (87,92.85) (88,93.05) (89,93.26) (90,93.45) (91,93.64) (92,93.83) (93,94.01) (94,94.18) (95,94.35) (96,94.51) (97,94.67) (98,94.83) (99,94.98) (100,95.12) (101,95.26) (102,95.40) (103,95.53) (104,95.66) (105,95.79) (106,95.91) (107,96.03) (108,96.15) (109,96.26) (110,96.37) (111,96.47) (112,96.57) (113,96.67) (114,96.77) (115,96.86) (116,96.95) (117,97.04) (118,97.13) (119,97.21) (120,97.29) (121,97.37) (122,97.45) (123,97.52) (124,97.59) (125,97.66) (126,97.73) (127,97.80) (128,97.86) (129,97.92) (130,97.98) (131,98.04) (132,98.10) (133,98.15) (134,98.21) (135,98.26) (136,98.31) (137,98.36) (138,98.41) (139,98.45) (140,98.50) (141,98.54) (142,98.58) (143,98.62) (144,98.66) (145,98.70) (146,98.74) (147,98.78) (148,98.81) (149,98.85) (150,98.88) (151,98.91) (152,98.95) (153,98.98) (154,99.01) (155,99.03) (156,99.06) (157,99.09) (158,99.12) (159,99.14) (160,99.17) (161,99.19) (162,99.21) (163,99.24) (164,99.26) (165,99.28) (166,99.30) (167,99.32) (168,99.34) (169,99.36) (170,99.38) (171,99.40) (172,99.41) (173,99.43) (174,99.45) (175,99.46) (176,99.48) (177,99.49) (178,99.51) (179,99.52) (180,99.54) (181,99.55) (182,99.56) (183,99.58) (184,99.59) (185,99.60) (186,99.61) (187,99.62) (188,99.63) (189,99.65) (190,99.66) (191,99.67) (192,99.68) (193,99.68) (194,99.69) (195,99.70) (196,99.71) (197,99.72) (198,99.73) (199,99.74) (200,99.74) (201,99.75) (202,99.76) (203,99.77) (204,99.77) (205,99.78) (206,99.78) (207,99.79) (208,99.80) (209,99.80) (210,99.81) (211,99.81) (212,99.82) (213,99.82) (214,99.83) (215,99.83) (216,99.84) (217,99.84) (218,99.85) (219,99.85) (220,99.86) (221,99.86) (222,99.87) (223,99.87) (224,99.87) (225,99.88) (226,99.88) (227,99.88) (228,99.89) (229,99.89) (230,99.89) (231,99.90) (232,99.90) (233,99.90) (234,99.91) (235,99.91) (236,99.91) (237,99.91) (238,99.92) (239,99.92) (240,99.92) (241,99.92) (242,99.93) (243,99.93) (244,99.93) (245,99.93) (246,99.93) (247,99.94) (248,99.94) (249,99.94) (250,99.94) (251,99.94) (252,99.94) (253,99.95) (254,99.95) (255,99.95) (256,99.95) (257,99.95) (258,99.95) (259,99.95) (260,99.96) (261,99.96) (262,99.96) (263,99.96) (264,99.96) (265,99.96) (266,99.96) (267,99.96) (268,99.97) (269,99.97) (270,99.97) (271,99.97) (272,99.97) (273,99.97) (274,99.97) (275,99.97) (276,99.97) (277,99.97) (278,99.97) (279,99.97) (280,99.98) (281,99.98) (282,99.98) (283,99.98) (284,99.98) (285,99.98) (286,99.98) (287,99.98) (288,99.98) (289,99.98) (290,99.98) (291,99.98) (292,99.98) (293,99.98) (294,99.98) (295,99.98) (296,99.98) (297,99.99) (298,99.99) (299,99.99) (300,99.99) (301,99.99) (302,99.99) (303,99.99) (304,99.99) (305,99.99) (306,99.99) (307,99.99) (308,99.99) (309,99.99) (310,99.99) (311,99.99) (312,99.99) (313,99.99) (314,99.99) (315,99.99) (316,99.99) (317,99.99) (318,99.99) (319,99.99) (320,99.99) (321,99.99) (322,99.99) (323,99.99) (324,99.99) (325,99.99) (326,99.99) (327,99.99) (328,99.99) (329,99.99) (330,99.99) (331,99.99) (332,99.99) (333,99.99) (334,100.00) (335,100.00) (336,100.00) (337,100.00) (338,100.00) (339,100.00) (340,100.00) (341,100.00) (342,100.00) (343,100.00) (344,100.00) (345,100.00) (346,100.00) (347,100.00) (348,100.00) (349,100.00) (350,100.00) (351,100.00) (352,100.00) (353,100.00) (354,100.00) (355,100.00) (356,100.00) (357,100.00) (358,100.00) (359,100.00) (360,100.00) (361,100.00) (362,100.00) (363,100.00) (364,100.00) (365,100.00) (366,100.00) (367,100.00) (368,100.00) (369,100.00) (370,100.00) (371,100.00) (372,100.00) (373,100.00) (374,100.00) (375,100.00) (376,100.00) (377,100.00) (378,100.00) (379,100.00) (380,100.00) (381,100.00) (382,100.00) (383,100.00) (384,100.00) (385,100.00) (386,100.00) (387,100.00) (388,100.00) (389,100.00) (390,100.00) (391,100.00) (392,100.00) (393,100.00) (394,100.00) (395,100.00) (396,100.00) (397,100.00) (398,100.00) (399,100.00)  }; \addlegendentry{exponential model}
    \addplot[smooth,color=red] plot coordinates { (1,10.00) (2, 6.25) (3, 5.00) (4, 4.38) (5, 4.00) (6, 3.75) (7, 3.57) (8, 3.44) (9, 3.33) (10, 3.25) (11, 3.18) (12, 3.12) (13, 3.08) (14, 3.04) (15, 3.00) (16, 2.97) (17, 2.94) (18, 2.92) (19, 2.89) (20, 2.88) (21, 2.86) (22, 2.84) (23, 2.83) (24, 2.81) (25, 2.80) (26, 2.79) (27, 2.78) (28, 2.77) (29, 2.76) (30, 2.75) (31, 2.74) (32, 2.73) (33, 2.73) (34, 2.72) (35, 2.71) (36, 2.71) (37, 2.70) (38, 2.63) (39, 2.56) (40, 2.50) (41, 2.44) (42, 2.38) (43, 2.33) (44, 2.27) (45, 2.22) (46, 2.17) (47, 2.13) (48, 2.08) (49, 2.04) (50, 2.00) (51, 1.96) (52, 1.92) (53, 1.89) (54, 1.85) (55, 1.82) (56, 1.79) (57, 1.75) (58, 1.72) (59, 1.69) (60, 1.67) (61, 1.64) (62, 1.61) (63, 1.59) (64, 1.56) (65, 1.54) (66, 1.52) (67, 1.49) (68, 1.47) (69, 1.45) (70, 1.43) (71, 1.41) (72, 1.39) (73, 1.37) (74, 1.35) (75, 1.33) (76, 1.32) (77, 1.30) (78, 1.28) (79, 1.27) (80, 1.25) (81, 1.23) (82, 1.22) (83, 1.20) (84, 1.19) (85, 1.18) (86, 1.16) (87, 1.15) (88, 1.14) (89, 1.12) (90, 1.11) (91, 1.10) (92, 1.09) (93, 1.08) (94, 1.06) (95, 1.05) (96, 1.04) (97, 1.03) (98, 1.02) (99, 1.01) (100, 1.00) (101, 0.99) (102, 0.98) (103, 0.97) (104, 0.96) (105, 0.95) (106, 0.94) (107, 0.93) (108, 0.93) (109, 0.92) (110, 0.91) (111, 0.90) (112, 0.89) (113, 0.88) (114, 0.88) (115, 0.87) (116, 0.86) (117, 0.85) (118, 0.85) (119, 0.84) (120, 0.83) (121, 0.83) (122, 0.82) (123, 0.81) (124, 0.81) (125, 0.80) (126, 0.79) (127, 0.79) (128, 0.78) (129, 0.78) (130, 0.77) (131, 0.76) (132, 0.76) (133, 0.75) (134, 0.75) (135, 0.74) (136, 0.74) (137, 0.73) (138, 0.72) (139, 0.72) (140, 0.71) (141, 0.71) (142, 0.70) (143, 0.70) (144, 0.69) (145, 0.69) (146, 0.68) (147, 0.68) (148, 0.68) (149, 0.67) (150, 0.67) (151, 0.66) (152, 0.66) (153, 0.65) (154, 0.65) (155, 0.65) (156, 0.64) (157, 0.64) (158, 0.63) (159, 0.63) (160, 0.62) (161, 0.62) (162, 0.62) (163, 0.61) (164, 0.61) (165, 0.61) (166, 0.60) (167, 0.60) (168, 0.60) (169, 0.59) (170, 0.59) (171, 0.58) (172, 0.58) (173, 0.58) (174, 0.57) (175, 0.57) (176, 0.57) (177, 0.56) (178, 0.56) (179, 0.56) (180, 0.56) (181, 0.55) (182, 0.55) (183, 0.55) (184, 0.54) (185, 0.54) (186, 0.54) (187, 0.53) (188, 0.53) (189, 0.53) (190, 0.53) (191, 0.52) (192, 0.52) (193, 0.52) (194, 0.52) (195, 0.51) (196, 0.51) (197, 0.51) (198, 0.51) (199, 0.50) (200, 0.50) (201, 0.50) (202, 0.50) (203, 0.49) (204, 0.49) (205, 0.49) (206, 0.49) (207, 0.48) (208, 0.48) (209, 0.48) (210, 0.48) (211, 0.47) (212, 0.47) (213, 0.47) (214, 0.47) (215, 0.47) (216, 0.46) (217, 0.46) (218, 0.46) (219, 0.46) (220, 0.45) (221, 0.45) (222, 0.45) (223, 0.45) (224, 0.45) (225, 0.44) (226, 0.44) (227, 0.44) (228, 0.44) (229, 0.44) (230, 0.43) (231, 0.43) (232, 0.43) (233, 0.43) (234, 0.43) (235, 0.43) (236, 0.42) (237, 0.42) (238, 0.42) (239, 0.42) (240, 0.42) (241, 0.41) (242, 0.41) (243, 0.41) (244, 0.41) (245, 0.41) (246, 0.41) (247, 0.40) (248, 0.40) (249, 0.40) (250, 0.40) (251, 0.40) (252, 0.40) (253, 0.40) (254, 0.39) (255, 0.39) (256, 0.39) (257, 0.39) (258, 0.39) (259, 0.39) (260, 0.38) (261, 0.38) (262, 0.38) (263, 0.38) (264, 0.38) (265, 0.38) (266, 0.38) (267, 0.37) (268, 0.37) (269, 0.37) (270, 0.37) (271, 0.37) (272, 0.37) (273, 0.37) (274, 0.36) (275, 0.36) (276, 0.36) (277, 0.36) (278, 0.36) (279, 0.36) (280, 0.36) (281, 0.36) (282, 0.35) (283, 0.35) (284, 0.35) (285, 0.35) (286, 0.35) (287, 0.35) (288, 0.35) (289, 0.35) (290, 0.34) (291, 0.34) (292, 0.34) (293, 0.34) (294, 0.34) (295, 0.34) (296, 0.34) (297, 0.34) (298, 0.34) (299, 0.33) (300, 0.33) (301, 0.33) (302, 0.33) (303, 0.33) (304, 0.33) (305, 0.33) (306, 0.33) (307, 0.33) (308, 0.32) (309, 0.32) (310, 0.32) (311, 0.32) (312, 0.32) (313, 0.32) (314, 0.32) (315, 0.32) (316, 0.32) (317, 0.32) (318, 0.31) (319, 0.31) (320, 0.31) (321, 0.31) (322, 0.31) (323, 0.31) (324, 0.31) (325, 0.31) (326, 0.31) (327, 0.31) (328, 0.30) (329, 0.30) (330, 0.30) (331, 0.30) (332, 0.30) (333, 0.30) (334, 0.30) (335, 0.30) (336, 0.30) (337, 0.30) (338, 0.30) (339, 0.29) (340, 0.29) (341, 0.29) (342, 0.29) (343, 0.29) (344, 0.29) (345, 0.29) (346, 0.29) (347, 0.29) (348, 0.29) (349, 0.29) (350, 0.29) (351, 0.28) (352, 0.28) (353, 0.28) (354, 0.28) (355, 0.28) (356, 0.28) (357, 0.28) (358, 0.28) (359, 0.28) (360, 0.28) (361, 0.28) (362, 0.28) (363, 0.28) (364, 0.27) (365, 0.27) (366, 0.27) (367, 0.27) (368, 0.27) (369, 0.27) (370, 0.27) (371, 0.27) (372, 0.27) (373, 0.27) (374, 0.27) (375, 0.27) (376, 0.27) (377, 0.27) (378, 0.26) (379, 0.26) (380, 0.26) (381, 0.26) (382, 0.26) (383, 0.26) (384, 0.26) (385, 0.26) (386, 0.26) (387, 0.26) (388, 0.26) (389, 0.26) (390, 0.26) (391, 0.26) (392, 0.26) (393, 0.25) (394, 0.25) (395, 0.25) (396, 0.25) (397, 0.25) (398, 0.25) (399, 0.25)  };
    \addplot[smooth,color=red, dashed] plot coordinates { (1,10.00) (2, 6.31) (3, 5.05) (4, 4.40) (5, 4.00) (6, 3.72) (7, 3.51) (8, 3.35) (9, 3.21) (10, 3.10) (11, 3.00) (12, 2.91) (13, 2.83) (14, 2.76) (15, 2.69) (16, 2.63) (17, 2.58) (18, 2.52) (19, 2.48) (20, 2.43) (21, 2.38) (22, 2.34) (23, 2.30) (24, 2.26) (25, 2.22) (26, 2.19) (27, 2.15) (28, 2.12) (29, 2.09) (30, 2.06) (31, 2.03) (32, 2.00) (33, 1.97) (34, 1.94) (35, 1.91) (36, 1.89) (37, 1.86) (38, 1.83) (39, 1.81) (40, 1.79) (41, 1.76) (42, 1.74) (43, 1.72) (44, 1.70) (45, 1.67) (46, 1.65) (47, 1.63) (48, 1.61) (49, 1.59) (50, 1.57) (51, 1.56) (52, 1.54) (53, 1.52) (54, 1.50) (55, 1.48) (56, 1.47) (57, 1.45) (58, 1.43) (59, 1.42) (60, 1.40) (61, 1.39) (62, 1.37) (63, 1.36) (64, 1.34) (65, 1.33) (66, 1.31) (67, 1.30) (68, 1.29) (69, 1.27) (70, 1.26) (71, 1.25) (72, 1.23) (73, 1.22) (74, 1.21) (75, 1.20) (76, 1.19) (77, 1.17) (78, 1.16) (79, 1.15) (80, 1.14) (81, 1.13) (82, 1.12) (83, 1.11) (84, 1.10) (85, 1.09) (86, 1.08) (87, 1.07) (88, 1.06) (89, 1.05) (90, 1.04) (91, 1.03) (92, 1.02) (93, 1.01) (94, 1.00) (95, 0.99) (96, 0.98) (97, 0.98) (98, 0.97) (99, 0.96) (100, 0.95) (101, 0.94) (102, 0.94) (103, 0.93) (104, 0.92) (105, 0.91) (106, 0.90) (107, 0.90) (108, 0.89) (109, 0.88) (110, 0.88) (111, 0.87) (112, 0.86) (113, 0.86) (114, 0.85) (115, 0.84) (116, 0.84) (117, 0.83) (118, 0.82) (119, 0.82) (120, 0.81) (121, 0.80) (122, 0.80) (123, 0.79) (124, 0.79) (125, 0.78) (126, 0.78) (127, 0.77) (128, 0.76) (129, 0.76) (130, 0.75) (131, 0.75) (132, 0.74) (133, 0.74) (134, 0.73) (135, 0.73) (136, 0.72) (137, 0.72) (138, 0.71) (139, 0.71) (140, 0.70) (141, 0.70) (142, 0.69) (143, 0.69) (144, 0.69) (145, 0.68) (146, 0.68) (147, 0.67) (148, 0.67) (149, 0.66) (150, 0.66) (151, 0.66) (152, 0.65) (153, 0.65) (154, 0.64) (155, 0.64) (156, 0.64) (157, 0.63) (158, 0.63) (159, 0.62) (160, 0.62) (161, 0.62) (162, 0.61) (163, 0.61) (164, 0.61) (165, 0.60) (166, 0.60) (167, 0.59) (168, 0.59) (169, 0.59) (170, 0.58) (171, 0.58) (172, 0.58) (173, 0.57) (174, 0.57) (175, 0.57) (176, 0.57) (177, 0.56) (178, 0.56) (179, 0.56) (180, 0.55) (181, 0.55) (182, 0.55) (183, 0.54) (184, 0.54) (185, 0.54) (186, 0.54) (187, 0.53) (188, 0.53) (189, 0.53) (190, 0.52) (191, 0.52) (192, 0.52) (193, 0.52) (194, 0.51) (195, 0.51) (196, 0.51) (197, 0.51) (198, 0.50) (199, 0.50) (200, 0.50) (201, 0.50) (202, 0.49) (203, 0.49) (204, 0.49) (205, 0.49) (206, 0.48) (207, 0.48) (208, 0.48) (209, 0.48) (210, 0.48) (211, 0.47) (212, 0.47) (213, 0.47) (214, 0.47) (215, 0.46) (216, 0.46) (217, 0.46) (218, 0.46) (219, 0.46) (220, 0.45) (221, 0.45) (222, 0.45) (223, 0.45) (224, 0.45) (225, 0.44) (226, 0.44) (227, 0.44) (228, 0.44) (229, 0.44) (230, 0.43) (231, 0.43) (232, 0.43) (233, 0.43) (234, 0.43) (235, 0.43) (236, 0.42) (237, 0.42) (238, 0.42) (239, 0.42) (240, 0.42) (241, 0.41) (242, 0.41) (243, 0.41) (244, 0.41) (245, 0.41) (246, 0.41) (247, 0.40) (248, 0.40) (249, 0.40) (250, 0.40) (251, 0.40) (252, 0.40) (253, 0.40) (254, 0.39) (255, 0.39) (256, 0.39) (257, 0.39) (258, 0.39) (259, 0.39) (260, 0.38) (261, 0.38) (262, 0.38) (263, 0.38) (264, 0.38) (265, 0.38) (266, 0.38) (267, 0.37) (268, 0.37) (269, 0.37) (270, 0.37) (271, 0.37) (272, 0.37) (273, 0.37) (274, 0.36) (275, 0.36) (276, 0.36) (277, 0.36) (278, 0.36) (279, 0.36) (280, 0.36) (281, 0.36) (282, 0.35) (283, 0.35) (284, 0.35) (285, 0.35) (286, 0.35) (287, 0.35) (288, 0.35) (289, 0.35) (290, 0.34) (291, 0.34) (292, 0.34) (293, 0.34) (294, 0.34) (295, 0.34) (296, 0.34) (297, 0.34) (298, 0.34) (299, 0.33) (300, 0.33) (301, 0.33) (302, 0.33) (303, 0.33) (304, 0.33) (305, 0.33) (306, 0.33) (307, 0.33) (308, 0.32) (309, 0.32) (310, 0.32) (311, 0.32) (312, 0.32) (313, 0.32) (314, 0.32) (315, 0.32) (316, 0.32) (317, 0.32) (318, 0.31) (319, 0.31) (320, 0.31) (321, 0.31) (322, 0.31) (323, 0.31) (324, 0.31) (325, 0.31) (326, 0.31) (327, 0.31) (328, 0.30) (329, 0.30) (330, 0.30) (331, 0.30) (332, 0.30) (333, 0.30) (334, 0.30) (335, 0.30) (336, 0.30) (337, 0.30) (338, 0.30) (339, 0.29) (340, 0.29) (341, 0.29) (342, 0.29) (343, 0.29) (344, 0.29) (345, 0.29) (346, 0.29) (347, 0.29) (348, 0.29) (349, 0.29) (350, 0.29) (351, 0.28) (352, 0.28) (353, 0.28) (354, 0.28) (355, 0.28) (356, 0.28) (357, 0.28) (358, 0.28) (359, 0.28) (360, 0.28) (361, 0.28) (362, 0.28) (363, 0.28) (364, 0.27) (365, 0.27) (366, 0.27) (367, 0.27) (368, 0.27) (369, 0.27) (370, 0.27) (371, 0.27) (372, 0.27) (373, 0.27) (374, 0.27) (375, 0.27) (376, 0.27) (377, 0.27) (378, 0.26) (379, 0.26) (380, 0.26) (381, 0.26) (382, 0.26) (383, 0.26) (384, 0.26) (385, 0.26) (386, 0.26) (387, 0.26) (388, 0.26) (389, 0.26) (390, 0.26) (391, 0.26) (392, 0.26) (393, 0.25) (394, 0.25) (395, 0.25) (396, 0.25) (397, 0.25) (398, 0.25) (399, 0.25)  };
    \addplot[color=gray,mark=*] plot coordinates { (1,10) };
  \end{axis}
\end{tikzpicture}
\caption{\it Les courbes de prix et de revenu de la même transaction $\textcolor{blue}{\boxed{{\bf 10}\sqcup^{\mbox{\tiny 100}}}}$, la même valeur de $\xi=0.25$, pour le modèle linéaire par morceaux et pour le modèle exponentiel}
\label{diff_model}
\end{figure}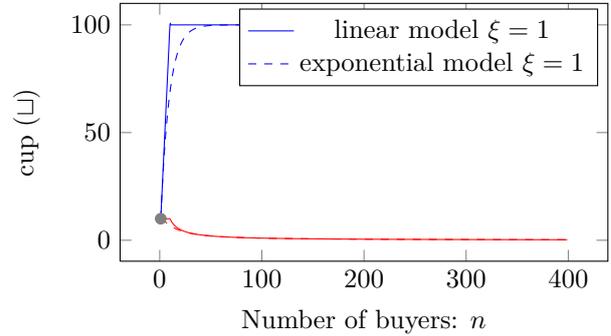
\begin{figure}
\begin{tikzpicture}
  \begin{axis}[ xlabel=Number of buyers: $n$, ylabel=cup ($\sqcup$), width=8cm, height=5cm]
    \addplot[smooth,blue] plot coordinates { (1,10.00) (2,20.00) (3,30.00) (4,40.00) (5,50.00) (6,60.00) (7,70.00) (8,80.00) (9,90.00) (10,100.00) (11,100.00) (12,100.00) (13,100.00) (14,100.00) (15,100.00) (16,100.00) (17,100.00) (18,100.00) (19,100.00) (20,100.00) (21,100.00) (22,100.00) (23,100.00) (24,100.00) (25,100.00) (26,100.00) (27,100.00) (28,100.00) (29,100.00) (30,100.00) (31,100.00) (32,100.00) (33,100.00) (34,100.00) (35,100.00) (36,100.00) (37,100.00) (38,100.00) (39,100.00) (40,100.00) (41,100.00) (42,100.00) (43,100.00) (44,100.00) (45,100.00) (46,100.00) (47,100.00) (48,100.00) (49,100.00) (50,100.00) (51,100.00) (52,100.00) (53,100.00) (54,100.00) (55,100.00) (56,100.00) (57,100.00) (58,100.00) (59,100.00) (60,100.00) (61,100.00) (62,100.00) (63,100.00) (64,100.00) (65,100.00) (66,100.00) (67,100.00) (68,100.00) (69,100.00) (70,100.00) (71,100.00) (72,100.00) (73,100.00) (74,100.00) (75,100.00) (76,100.00) (77,100.00) (78,100.00) (79,100.00) (80,100.00) (81,100.00) (82,100.00) (83,100.00) (84,100.00) (85,100.00) (86,100.00) (87,100.00) (88,100.00) (89,100.00) (90,100.00) (91,100.00) (92,100.00) (93,100.00) (94,100.00) (95,100.00) (96,100.00) (97,100.00) (98,100.00) (99,100.00) (100,100.00) (101,100.00) (102,100.00) (103,100.00) (104,100.00) (105,100.00) (106,100.00) (107,100.00) (108,100.00) (109,100.00) (110,100.00) (111,100.00) (112,100.00) (113,100.00) (114,100.00) (115,100.00) (116,100.00) (117,100.00) (118,100.00) (119,100.00) (120,100.00) (121,100.00) (122,100.00) (123,100.00) (124,100.00) (125,100.00) (126,100.00) (127,100.00) (128,100.00) (129,100.00) (130,100.00) (131,100.00) (132,100.00) (133,100.00) (134,100.00) (135,100.00) (136,100.00) (137,100.00) (138,100.00) (139,100.00) (140,100.00) (141,100.00) (142,100.00) (143,100.00) (144,100.00) (145,100.00) (146,100.00) (147,100.00) (148,100.00) (149,100.00) (150,100.00) (151,100.00) (152,100.00) (153,100.00) (154,100.00) (155,100.00) (156,100.00) (157,100.00) (158,100.00) (159,100.00) (160,100.00) (161,100.00) (162,100.00) (163,100.00) (164,100.00) (165,100.00) (166,100.00) (167,100.00) (168,100.00) (169,100.00) (170,100.00) (171,100.00) (172,100.00) (173,100.00) (174,100.00) (175,100.00) (176,100.00) (177,100.00) (178,100.00) (179,100.00) (180,100.00) (181,100.00) (182,100.00) (183,100.00) (184,100.00) (185,100.00) (186,100.00) (187,100.00) (188,100.00) (189,100.00) (190,100.00) (191,100.00) (192,100.00) (193,100.00) (194,100.00) (195,100.00) (196,100.00) (197,100.00) (198,100.00) (199,100.00) (200,100.00) (201,100.00) (202,100.00) (203,100.00) (204,100.00) (205,100.00) (206,100.00) (207,100.00) (208,100.00) (209,100.00) (210,100.00) (211,100.00) (212,100.00) (213,100.00) (214,100.00) (215,100.00) (216,100.00) (217,100.00) (218,100.00) (219,100.00) (220,100.00) (221,100.00) (222,100.00) (223,100.00) (224,100.00) (225,100.00) (226,100.00) (227,100.00) (228,100.00) (229,100.00) (230,100.00) (231,100.00) (232,100.00) (233,100.00) (234,100.00) (235,100.00) (236,100.00) (237,100.00) (238,100.00) (239,100.00) (240,100.00) (241,100.00) (242,100.00) (243,100.00) (244,100.00) (245,100.00) (246,100.00) (247,100.00) (248,100.00) (249,100.00) (250,100.00) (251,100.00) (252,100.00) (253,100.00) (254,100.00) (255,100.00) (256,100.00) (257,100.00) (258,100.00) (259,100.00) (260,100.00) (261,100.00) (262,100.00) (263,100.00) (264,100.00) (265,100.00) (266,100.00) (267,100.00) (268,100.00) (269,100.00) (270,100.00) (271,100.00) (272,100.00) (273,100.00) (274,100.00) (275,100.00) (276,100.00) (277,100.00) (278,100.00) (279,100.00) (280,100.00) (281,100.00) (282,100.00) (283,100.00) (284,100.00) (285,100.00) (286,100.00) (287,100.00) (288,100.00) (289,100.00) (290,100.00) (291,100.00) (292,100.00) (293,100.00) (294,100.00) (295,100.00) (296,100.00) (297,100.00) (298,100.00) (299,100.00) (300,100.00) (301,100.00) (302,100.00) (303,100.00) (304,100.00) (305,100.00) (306,100.00) (307,100.00) (308,100.00) (309,100.00) (310,100.00) (311,100.00) (312,100.00) (313,100.00) (314,100.00) (315,100.00) (316,100.00) (317,100.00) (318,100.00) (319,100.00) (320,100.00) (321,100.00) (322,100.00) (323,100.00) (324,100.00) (325,100.00) (326,100.00) (327,100.00) (328,100.00) (329,100.00) (330,100.00) (331,100.00) (332,100.00) (333,100.00) (334,100.00) (335,100.00) (336,100.00) (337,100.00) (338,100.00) (339,100.00) (340,100.00) (341,100.00) (342,100.00) (343,100.00) (344,100.00) (345,100.00) (346,100.00) (347,100.00) (348,100.00) (349,100.00) (350,100.00) (351,100.00) (352,100.00) (353,100.00) (354,100.00) (355,100.00) (356,100.00) (357,100.00) (358,100.00) (359,100.00) (360,100.00) (361,100.00) (362,100.00) (363,100.00) (364,100.00) (365,100.00) (366,100.00) (367,100.00) (368,100.00) (369,100.00) (370,100.00) (371,100.00) (372,100.00) (373,100.00) (374,100.00) (375,100.00) (376,100.00) (377,100.00) (378,100.00) (379,100.00) (380,100.00) (381,100.00) (382,100.00) (383,100.00) (384,100.00) (385,100.00) (386,100.00) (387,100.00) (388,100.00) (389,100.00) (390,100.00) (391,100.00) (392,100.00) (393,100.00) (394,100.00) (395,100.00) (396,100.00) (397,100.00) (398,100.00) (399,100.00)  };\addlegendentry{linear model $\xi=1$}
    \addplot[smooth,blue, dashed] plot coordinates { (1,10.00) (2,20.00) (3,28.89) (4,36.79) (5,43.81) (6,50.06) (7,55.61) (8,60.54) (9,64.92) (10,68.82) (11,72.28) (12,75.36) (13,78.10) (14,80.53) (15,82.70) (16,84.62) (17,86.33) (18,87.85) (19,89.20) (20,90.40) (21,91.47) (22,92.41) (23,93.26) (24,94.01) (25,94.67) (26,95.26) (27,95.79) (28,96.26) (29,96.67) (30,97.04) (31,97.37) (32,97.66) (33,97.92) (34,98.15) (35,98.36) (36,98.54) (37,98.70) (38,98.85) (39,98.98) (40,99.09) (41,99.19) (42,99.28) (43,99.36) (44,99.43) (45,99.49) (46,99.55) (47,99.60) (48,99.65) (49,99.68) (50,99.72) (51,99.75) (52,99.78) (53,99.80) (54,99.82) (55,99.84) (56,99.86) (57,99.88) (58,99.89) (59,99.90) (60,99.91) (61,99.92) (62,99.93) (63,99.94) (64,99.95) (65,99.95) (66,99.96) (67,99.96) (68,99.97) (69,99.97) (70,99.97) (71,99.98) (72,99.98) (73,99.98) (74,99.98) (75,99.99) (76,99.99) (77,99.99) (78,99.99) (79,99.99) (80,99.99) (81,99.99) (82,99.99) (83,99.99) (84,99.99) (85,100.00) (86,100.00) (87,100.00) (88,100.00) (89,100.00) (90,100.00) (91,100.00) (92,100.00) (93,100.00) (94,100.00) (95,100.00) (96,100.00) (97,100.00) (98,100.00) (99,100.00) (100,100.00) (101,100.00) (102,100.00) (103,100.00) (104,100.00) (105,100.00) (106,100.00) (107,100.00) (108,100.00) (109,100.00) (110,100.00) (111,100.00) (112,100.00) (113,100.00) (114,100.00) (115,100.00) (116,100.00) (117,100.00) (118,100.00) (119,100.00) (120,100.00) (121,100.00) (122,100.00) (123,100.00) (124,100.00) (125,100.00) (126,100.00) (127,100.00) (128,100.00) (129,100.00) (130,100.00) (131,100.00) (132,100.00) (133,100.00) (134,100.00) (135,100.00) (136,100.00) (137,100.00) (138,100.00) (139,100.00) (140,100.00) (141,100.00) (142,100.00) (143,100.00) (144,100.00) (145,100.00) (146,100.00) (147,100.00) (148,100.00) (149,100.00) (150,100.00) (151,100.00) (152,100.00) (153,100.00) (154,100.00) (155,100.00) (156,100.00) (157,100.00) (158,100.00) (159,100.00) (160,100.00) (161,100.00) (162,100.00) (163,100.00) (164,100.00) (165,100.00) (166,100.00) (167,100.00) (168,100.00) (169,100.00) (170,100.00) (171,100.00) (172,100.00) (173,100.00) (174,100.00) (175,100.00) (176,100.00) (177,100.00) (178,100.00) (179,100.00) (180,100.00) (181,100.00) (182,100.00) (183,100.00) (184,100.00) (185,100.00) (186,100.00) (187,100.00) (188,100.00) (189,100.00) (190,100.00) (191,100.00) (192,100.00) (193,100.00) (194,100.00) (195,100.00) (196,100.00) (197,100.00) (198,100.00) (199,100.00) (200,100.00) (201,100.00) (202,100.00) (203,100.00) (204,100.00) (205,100.00) (206,100.00) (207,100.00) (208,100.00) (209,100.00) (210,100.00) (211,100.00) (212,100.00) (213,100.00) (214,100.00) (215,100.00) (216,100.00) (217,100.00) (218,100.00) (219,100.00) (220,100.00) (221,100.00) (222,100.00) (223,100.00) (224,100.00) (225,100.00) (226,100.00) (227,100.00) (228,100.00) (229,100.00) (230,100.00) (231,100.00) (232,100.00) (233,100.00) (234,100.00) (235,100.00) (236,100.00) (237,100.00) (238,100.00) (239,100.00) (240,100.00) (241,100.00) (242,100.00) (243,100.00) (244,100.00) (245,100.00) (246,100.00) (247,100.00) (248,100.00) (249,100.00) (250,100.00) (251,100.00) (252,100.00) (253,100.00) (254,100.00) (255,100.00) (256,100.00) (257,100.00) (258,100.00) (259,100.00) (260,100.00) (261,100.00) (262,100.00) (263,100.00) (264,100.00) (265,100.00) (266,100.00) (267,100.00) (268,100.00) (269,100.00) (270,100.00) (271,100.00) (272,100.00) (273,100.00) (274,100.00) (275,100.00) (276,100.00) (277,100.00) (278,100.00) (279,100.00) (280,100.00) (281,100.00) (282,100.00) (283,100.00) (284,100.00) (285,100.00) (286,100.00) (287,100.00) (288,100.00) (289,100.00) (290,100.00) (291,100.00) (292,100.00) (293,100.00) (294,100.00) (295,100.00) (296,100.00) (297,100.00) (298,100.00) (299,100.00) (300,100.00) (301,100.00) (302,100.00) (303,100.00) (304,100.00) (305,100.00) (306,100.00) (307,100.00) (308,100.00) (309,100.00) (310,100.00) (311,100.00) (312,100.00) (313,100.00) (314,100.00) (315,100.00) (316,100.00) (317,100.00) (318,100.00) (319,100.00) (320,100.00) (321,100.00) (322,100.00) (323,100.00) (324,100.00) (325,100.00) (326,100.00) (327,100.00) (328,100.00) (329,100.00) (330,100.00) (331,100.00) (332,100.00) (333,100.00) (334,100.00) (335,100.00) (336,100.00) (337,100.00) (338,100.00) (339,100.00) (340,100.00) (341,100.00) (342,100.00) (343,100.00) (344,100.00) (345,100.00) (346,100.00) (347,100.00) (348,100.00) (349,100.00) (350,100.00) (351,100.00) (352,100.00) (353,100.00) (354,100.00) (355,100.00) (356,100.00) (357,100.00) (358,100.00) (359,100.00) (360,100.00) (361,100.00) (362,100.00) (363,100.00) (364,100.00) (365,100.00) (366,100.00) (367,100.00) (368,100.00) (369,100.00) (370,100.00) (371,100.00) (372,100.00) (373,100.00) (374,100.00) (375,100.00) (376,100.00) (377,100.00) (378,100.00) (379,100.00) (380,100.00) (381,100.00) (382,100.00) (383,100.00) (384,100.00) (385,100.00) (386,100.00) (387,100.00) (388,100.00) (389,100.00) (390,100.00) (391,100.00) (392,100.00) (393,100.00) (394,100.00) (395,100.00) (396,100.00) (397,100.00) (398,100.00) (399,100.00)  }; \addlegendentry{exponential model $\xi=1$}
    \addplot[smooth,color=red] plot coordinates { (1,10.00) (2,10.00) (3,10.00) (4,10.00) (5,10.00) (6,10.00) (7,10.00) (8,10.00) (9,10.00) (10,10.00) (11, 9.09) (12, 8.33) (13, 7.69) (14, 7.14) (15, 6.67) (16, 6.25) (17, 5.88) (18, 5.56) (19, 5.26) (20, 5.00) (21, 4.76) (22, 4.55) (23, 4.35) (24, 4.17) (25, 4.00) (26, 3.85) (27, 3.70) (28, 3.57) (29, 3.45) (30, 3.33) (31, 3.23) (32, 3.12) (33, 3.03) (34, 2.94) (35, 2.86) (36, 2.78) (37, 2.70) (38, 2.63) (39, 2.56) (40, 2.50) (41, 2.44) (42, 2.38) (43, 2.33) (44, 2.27) (45, 2.22) (46, 2.17) (47, 2.13) (48, 2.08) (49, 2.04) (50, 2.00) (51, 1.96) (52, 1.92) (53, 1.89) (54, 1.85) (55, 1.82) (56, 1.79) (57, 1.75) (58, 1.72) (59, 1.69) (60, 1.67) (61, 1.64) (62, 1.61) (63, 1.59) (64, 1.56) (65, 1.54) (66, 1.52) (67, 1.49) (68, 1.47) (69, 1.45) (70, 1.43) (71, 1.41) (72, 1.39) (73, 1.37) (74, 1.35) (75, 1.33) (76, 1.32) (77, 1.30) (78, 1.28) (79, 1.27) (80, 1.25) (81, 1.23) (82, 1.22) (83, 1.20) (84, 1.19) (85, 1.18) (86, 1.16) (87, 1.15) (88, 1.14) (89, 1.12) (90, 1.11) (91, 1.10) (92, 1.09) (93, 1.08) (94, 1.06) (95, 1.05) (96, 1.04) (97, 1.03) (98, 1.02) (99, 1.01) (100, 1.00) (101, 0.99) (102, 0.98) (103, 0.97) (104, 0.96) (105, 0.95) (106, 0.94) (107, 0.93) (108, 0.93) (109, 0.92) (110, 0.91) (111, 0.90) (112, 0.89) (113, 0.88) (114, 0.88) (115, 0.87) (116, 0.86) (117, 0.85) (118, 0.85) (119, 0.84) (120, 0.83) (121, 0.83) (122, 0.82) (123, 0.81) (124, 0.81) (125, 0.80) (126, 0.79) (127, 0.79) (128, 0.78) (129, 0.78) (130, 0.77) (131, 0.76) (132, 0.76) (133, 0.75) (134, 0.75) (135, 0.74) (136, 0.74) (137, 0.73) (138, 0.72) (139, 0.72) (140, 0.71) (141, 0.71) (142, 0.70) (143, 0.70) (144, 0.69) (145, 0.69) (146, 0.68) (147, 0.68) (148, 0.68) (149, 0.67) (150, 0.67) (151, 0.66) (152, 0.66) (153, 0.65) (154, 0.65) (155, 0.65) (156, 0.64) (157, 0.64) (158, 0.63) (159, 0.63) (160, 0.62) (161, 0.62) (162, 0.62) (163, 0.61) (164, 0.61) (165, 0.61) (166, 0.60) (167, 0.60) (168, 0.60) (169, 0.59) (170, 0.59) (171, 0.58) (172, 0.58) (173, 0.58) (174, 0.57) (175, 0.57) (176, 0.57) (177, 0.56) (178, 0.56) (179, 0.56) (180, 0.56) (181, 0.55) (182, 0.55) (183, 0.55) (184, 0.54) (185, 0.54) (186, 0.54) (187, 0.53) (188, 0.53) (189, 0.53) (190, 0.53) (191, 0.52) (192, 0.52) (193, 0.52) (194, 0.52) (195, 0.51) (196, 0.51) (197, 0.51) (198, 0.51) (199, 0.50) (200, 0.50) (201, 0.50) (202, 0.50) (203, 0.49) (204, 0.49) (205, 0.49) (206, 0.49) (207, 0.48) (208, 0.48) (209, 0.48) (210, 0.48) (211, 0.47) (212, 0.47) (213, 0.47) (214, 0.47) (215, 0.47) (216, 0.46) (217, 0.46) (218, 0.46) (219, 0.46) (220, 0.45) (221, 0.45) (222, 0.45) (223, 0.45) (224, 0.45) (225, 0.44) (226, 0.44) (227, 0.44) (228, 0.44) (229, 0.44) (230, 0.43) (231, 0.43) (232, 0.43) (233, 0.43) (234, 0.43) (235, 0.43) (236, 0.42) (237, 0.42) (238, 0.42) (239, 0.42) (240, 0.42) (241, 0.41) (242, 0.41) (243, 0.41) (244, 0.41) (245, 0.41) (246, 0.41) (247, 0.40) (248, 0.40) (249, 0.40) (250, 0.40) (251, 0.40) (252, 0.40) (253, 0.40) (254, 0.39) (255, 0.39) (256, 0.39) (257, 0.39) (258, 0.39) (259, 0.39) (260, 0.38) (261, 0.38) (262, 0.38) (263, 0.38) (264, 0.38) (265, 0.38) (266, 0.38) (267, 0.37) (268, 0.37) (269, 0.37) (270, 0.37) (271, 0.37) (272, 0.37) (273, 0.37) (274, 0.36) (275, 0.36) (276, 0.36) (277, 0.36) (278, 0.36) (279, 0.36) (280, 0.36) (281, 0.36) (282, 0.35) (283, 0.35) (284, 0.35) (285, 0.35) (286, 0.35) (287, 0.35) (288, 0.35) (289, 0.35) (290, 0.34) (291, 0.34) (292, 0.34) (293, 0.34) (294, 0.34) (295, 0.34) (296, 0.34) (297, 0.34) (298, 0.34) (299, 0.33) (300, 0.33) (301, 0.33) (302, 0.33) (303, 0.33) (304, 0.33) (305, 0.33) (306, 0.33) (307, 0.33) (308, 0.32) (309, 0.32) (310, 0.32) (311, 0.32) (312, 0.32) (313, 0.32) (314, 0.32) (315, 0.32) (316, 0.32) (317, 0.32) (318, 0.31) (319, 0.31) (320, 0.31) (321, 0.31) (322, 0.31) (323, 0.31) (324, 0.31) (325, 0.31) (326, 0.31) (327, 0.31) (328, 0.30) (329, 0.30) (330, 0.30) (331, 0.30) (332, 0.30) (333, 0.30) (334, 0.30) (335, 0.30) (336, 0.30) (337, 0.30) (338, 0.30) (339, 0.29) (340, 0.29) (341, 0.29) (342, 0.29) (343, 0.29) (344, 0.29) (345, 0.29) (346, 0.29) (347, 0.29) (348, 0.29) (349, 0.29) (350, 0.29) (351, 0.28) (352, 0.28) (353, 0.28) (354, 0.28) (355, 0.28) (356, 0.28) (357, 0.28) (358, 0.28) (359, 0.28) (360, 0.28) (361, 0.28) (362, 0.28) (363, 0.28) (364, 0.27) (365, 0.27) (366, 0.27) (367, 0.27) (368, 0.27) (369, 0.27) (370, 0.27) (371, 0.27) (372, 0.27) (373, 0.27) (374, 0.27) (375, 0.27) (376, 0.27) (377, 0.27) (378, 0.26) (379, 0.26) (380, 0.26) (381, 0.26) (382, 0.26) (383, 0.26) (384, 0.26) (385, 0.26) (386, 0.26) (387, 0.26) (388, 0.26) (389, 0.26) (390, 0.26) (391, 0.26) (392, 0.26) (393, 0.25) (394, 0.25) (395, 0.25) (396, 0.25) (397, 0.25) (398, 0.25) (399, 0.25)  };
    \addplot[smooth,color=red, dashed] plot coordinates { (1,10.00) (2,10.00) (3, 9.63) (4, 9.20) (5, 8.76) (6, 8.34) (7, 7.94) (8, 7.57) (9, 7.21) (10, 6.88) (11, 6.57) (12, 6.28) (13, 6.01) (14, 5.75) (15, 5.51) (16, 5.29) (17, 5.08) (18, 4.88) (19, 4.69) (20, 4.52) (21, 4.36) (22, 4.20) (23, 4.05) (24, 3.92) (25, 3.79) (26, 3.66) (27, 3.55) (28, 3.44) (29, 3.33) (30, 3.23) (31, 3.14) (32, 3.05) (33, 2.97) (34, 2.89) (35, 2.81) (36, 2.74) (37, 2.67) (38, 2.60) (39, 2.54) (40, 2.48) (41, 2.42) (42, 2.36) (43, 2.31) (44, 2.26) (45, 2.21) (46, 2.16) (47, 2.12) (48, 2.08) (49, 2.03) (50, 1.99) (51, 1.96) (52, 1.92) (53, 1.88) (54, 1.85) (55, 1.82) (56, 1.78) (57, 1.75) (58, 1.72) (59, 1.69) (60, 1.67) (61, 1.64) (62, 1.61) (63, 1.59) (64, 1.56) (65, 1.54) (66, 1.51) (67, 1.49) (68, 1.47) (69, 1.45) (70, 1.43) (71, 1.41) (72, 1.39) (73, 1.37) (74, 1.35) (75, 1.33) (76, 1.32) (77, 1.30) (78, 1.28) (79, 1.27) (80, 1.25) (81, 1.23) (82, 1.22) (83, 1.20) (84, 1.19) (85, 1.18) (86, 1.16) (87, 1.15) (88, 1.14) (89, 1.12) (90, 1.11) (91, 1.10) (92, 1.09) (93, 1.08) (94, 1.06) (95, 1.05) (96, 1.04) (97, 1.03) (98, 1.02) (99, 1.01) (100, 1.00) (101, 0.99) (102, 0.98) (103, 0.97) (104, 0.96) (105, 0.95) (106, 0.94) (107, 0.93) (108, 0.93) (109, 0.92) (110, 0.91) (111, 0.90) (112, 0.89) (113, 0.88) (114, 0.88) (115, 0.87) (116, 0.86) (117, 0.85) (118, 0.85) (119, 0.84) (120, 0.83) (121, 0.83) (122, 0.82) (123, 0.81) (124, 0.81) (125, 0.80) (126, 0.79) (127, 0.79) (128, 0.78) (129, 0.78) (130, 0.77) (131, 0.76) (132, 0.76) (133, 0.75) (134, 0.75) (135, 0.74) (136, 0.74) (137, 0.73) (138, 0.72) (139, 0.72) (140, 0.71) (141, 0.71) (142, 0.70) (143, 0.70) (144, 0.69) (145, 0.69) (146, 0.68) (147, 0.68) (148, 0.68) (149, 0.67) (150, 0.67) (151, 0.66) (152, 0.66) (153, 0.65) (154, 0.65) (155, 0.65) (156, 0.64) (157, 0.64) (158, 0.63) (159, 0.63) (160, 0.62) (161, 0.62) (162, 0.62) (163, 0.61) (164, 0.61) (165, 0.61) (166, 0.60) (167, 0.60) (168, 0.60) (169, 0.59) (170, 0.59) (171, 0.58) (172, 0.58) (173, 0.58) (174, 0.57) (175, 0.57) (176, 0.57) (177, 0.56) (178, 0.56) (179, 0.56) (180, 0.56) (181, 0.55) (182, 0.55) (183, 0.55) (184, 0.54) (185, 0.54) (186, 0.54) (187, 0.53) (188, 0.53) (189, 0.53) (190, 0.53) (191, 0.52) (192, 0.52) (193, 0.52) (194, 0.52) (195, 0.51) (196, 0.51) (197, 0.51) (198, 0.51) (199, 0.50) (200, 0.50) (201, 0.50) (202, 0.50) (203, 0.49) (204, 0.49) (205, 0.49) (206, 0.49) (207, 0.48) (208, 0.48) (209, 0.48) (210, 0.48) (211, 0.47) (212, 0.47) (213, 0.47) (214, 0.47) (215, 0.47) (216, 0.46) (217, 0.46) (218, 0.46) (219, 0.46) (220, 0.45) (221, 0.45) (222, 0.45) (223, 0.45) (224, 0.45) (225, 0.44) (226, 0.44) (227, 0.44) (228, 0.44) (229, 0.44) (230, 0.43) (231, 0.43) (232, 0.43) (233, 0.43) (234, 0.43) (235, 0.43) (236, 0.42) (237, 0.42) (238, 0.42) (239, 0.42) (240, 0.42) (241, 0.41) (242, 0.41) (243, 0.41) (244, 0.41) (245, 0.41) (246, 0.41) (247, 0.40) (248, 0.40) (249, 0.40) (250, 0.40) (251, 0.40) (252, 0.40) (253, 0.40) (254, 0.39) (255, 0.39) (256, 0.39) (257, 0.39) (258, 0.39) (259, 0.39) (260, 0.38) (261, 0.38) (262, 0.38) (263, 0.38) (264, 0.38) (265, 0.38) (266, 0.38) (267, 0.37) (268, 0.37) (269, 0.37) (270, 0.37) (271, 0.37) (272, 0.37) (273, 0.37) (274, 0.36) (275, 0.36) (276, 0.36) (277, 0.36) (278, 0.36) (279, 0.36) (280, 0.36) (281, 0.36) (282, 0.35) (283, 0.35) (284, 0.35) (285, 0.35) (286, 0.35) (287, 0.35) (288, 0.35) (289, 0.35) (290, 0.34) (291, 0.34) (292, 0.34) (293, 0.34) (294, 0.34) (295, 0.34) (296, 0.34) (297, 0.34) (298, 0.34) (299, 0.33) (300, 0.33) (301, 0.33) (302, 0.33) (303, 0.33) (304, 0.33) (305, 0.33) (306, 0.33) (307, 0.33) (308, 0.32) (309, 0.32) (310, 0.32) (311, 0.32) (312, 0.32) (313, 0.32) (314, 0.32) (315, 0.32) (316, 0.32) (317, 0.32) (318, 0.31) (319, 0.31) (320, 0.31) (321, 0.31) (322, 0.31) (323, 0.31) (324, 0.31) (325, 0.31) (326, 0.31) (327, 0.31) (328, 0.30) (329, 0.30) (330, 0.30) (331, 0.30) (332, 0.30) (333, 0.30) (334, 0.30) (335, 0.30) (336, 0.30) (337, 0.30) (338, 0.30) (339, 0.29) (340, 0.29) (341, 0.29) (342, 0.29) (343, 0.29) (344, 0.29) (345, 0.29) (346, 0.29) (347, 0.29) (348, 0.29) (349, 0.29) (350, 0.29) (351, 0.28) (352, 0.28) (353, 0.28) (354, 0.28) (355, 0.28) (356, 0.28) (357, 0.28) (358, 0.28) (359, 0.28) (360, 0.28) (361, 0.28) (362, 0.28) (363, 0.28) (364, 0.27) (365, 0.27) (366, 0.27) (367, 0.27) (368, 0.27) (369, 0.27) (370, 0.27) (371, 0.27) (372, 0.27) (373, 0.27) (374, 0.27) (375, 0.27) (376, 0.27) (377, 0.27) (378, 0.26) (379, 0.26) (380, 0.26) (381, 0.26) (382, 0.26) (383, 0.26) (384, 0.26) (385, 0.26) (386, 0.26) (387, 0.26) (388, 0.26) (389, 0.26) (390, 0.26) (391, 0.26) (392, 0.26) (393, 0.25) (394, 0.25) (395, 0.25) (396, 0.25) (397, 0.25) (398, 0.25) (399, 0.25)  };
    \addplot[color=gray,mark=*] plot coordinates { (1,10) };
  \end{axis}
\end{tikzpicture}
\caption{\it Les courbes de prix et de revenu de la même transaction $\textcolor{blue}{\boxed{{\bf 10}\sqcup^{\mbox{\tiny 100}}}}$, avec $\xi=1$, pour le modèle linéaire par morceaux et pour le modèle exponentiel}
\label{diff_modelxi1}
\end{figure}\begin{table}
\begin{tabular}{|p{2.1cm}|p{2.4cm}|p{2.4cm}|}
\hline
{\bf Bien} &{\bf Matériel} & {\bf Immatériel} \\ \hline
acronyme & {\sc tg} & {\sc ig} \\ \hline
coût marginal & non nul & nul \\ \hline
rivalité & rival & non rival \\ \hline
identité & physique & numérique \\ \hline
propriété & dépossession & partage \\ \hline
type relation & 1-1 & 1-n \\ \hline
durée & instantanée & infinie \\ \hline
complexité & $\mathcal{O}(n)$ & $\mathcal{O}(n^2)$ \\ \hline
monnaie & région, état & monde \\ \hline
unité & 1 nombre & 2 nombres \\ \hline
spéculation & sensible & insensible \\ \hline
type marché & revente & vente directe \\ \hline
occasion & possible & impossible \\ \hline
lieux & magasins & réseau \net \\ \hline
intermédiation & oui & non \\ \hline
taxation & TVA & différence taux \\ \hline
naissance & 680 Av J.C. & 2015-2020 \\ \hline
\end{tabular}
\caption{différence entre le modèle économique des biens matériels ({\sc tg}) et celui des biens immatériels ({\sc ig}) }
\label{table_diff}
\end{table}
\section{Dans le monde réel}
Nous avons vu précédemment que la disponibilité partout et tout le temps d'un {\sc ig} acheté étaient une exigence légitime. Elle est réalisable à deux conditions~:
\begin{itemize}
\item La première est que les {\sc ig}s soient stockés (avec une certaine redondance) sur un réseau qui n'est pas effaçable. Cela implique une architecture {\em pair-à-pair} {\sc p2p}, utilisant les différentes mémoires des n\oe{}uds du réseau {\em Internet} et un nouveau protocole au dessus d'{\sc ip}.
 \item La deuxième condition est d'associer formellement l'{\sc ig} aux identités des humains qui l'entourent, c'est-à-dire, de son concepteur et de ses acheteurs. Cela passe par une identification et authentification forte sur tous les appareils permettant de lire l'{\sc ig}. Un avantage dérivé est que l'achat de l'{\sc ig} peut être compris dans l'achat de biens matériels ou de services, loisirs. Ainsi, une place de cinéma vous donne droit de revoir votre film sur téléphone, tablette ou {\sc tv} personnelle, mais il ne sera pas possible pour une autre personne de vous prendre ce même film. Le réseau détecterait que cette personne n'a pas acquitté son droit de lecture et donc ne peut avoir accès au film. De plus, vous pourrez revoir ce film toute votre vie. L'identité d'une personne est donc associée à un ensemble d'{\sc ig}s, correspondant à tous ses achats personnels. Une sorte de bibliothèque personnelle et confidentielle est donc stockée avec chiffrement sur {\em Internet}, véritable mémoire des biens culturels consommés ou créés.
On appelera {\em cerveau numérique} ({\em Digital Brain}) cette mémoire très personnelle.  
\end{itemize}

Les deux précédentes conditions sont aussi liées à des conditions très rigoureuses de sécurité et d'anonymat.

Divers outils autour d'{\em Internet} ont montré que l'on pouvait définir des protocoles décentralisés, sans autorité pour gérer le réseau. Ce point est plus difficile à admettre par certains qui voient en tout temps et tout lieux des pyramides décisionnelles. Par exemple, un administrateur {\em Unix} peut vous créer un compte, protégé par le mot de passe que vous aurez choisi seul, mais cet administrateur n'aura pas accès à ce mot de passe. Il pourra éventuellement le détruire. Or au sein d'un réseau {\sc p2p} sur {\em Internet}, il ne pourait même pas détruire ce mot de passe.

Il n'y a donc pas dans notre système de {\em BigBrother} contrôlant et suivant chaque transaction. 
Comme tout intervenant est identifié sur chaque appareil, il est possible d'utiliser les outils cryptographiques pour suivre les actions de cette personne tout en respectant ses données privées.

D'une part, tous les contenus représentant les {\sc ig} sont chiffrés, ils sont aussi signés (signature électronique) et horodatés\footnote{des serveurs dédiés de "{\em Time Stamping}" devront être installés en fonction du nombre important de requêtes. }.

Pour que le système remporte l'adhésion, il ne doit pas y avoir de "passe partout" ou de "méta clé" qui permettrait à une poignée d'individu d'avoir accès à cette mémoire commune. Chaque n\oe{}ud connecté à {\em Internet} met à disposition une partie de son processeur et une partie de sa mémoire, mais le contenu n'est jamais accessible, même à des fins de force majeure ou d'anti-terrorisme. Une taille suffisante de mémoire assure seulement que globalement, le système peut conserver les {\sc ig}s à plusieurs endroits si des n\oe{}uds venaient à être défaillants ou venaient à se déconnecter.

Prioritairement, le réseau sauvegarde sur le n\oe{}ud utilisé par une personne donnée, sa production en {\sc ig}, son historique de navigation et ses centres d'intérêts, dans une mémoire cache, de telle façon qu'en consultation hors ligne, l'utilisateur puisse éditer ou naviguer sur un espace le plus large possible. Il s'agit d'un principe bien connu en architecture des ordinateurs pour lesquels on recherche un étagement optimal de toutes les mémoires, depuis les registres du micro-processeur jusqu'à {\em Internet} en passant par la {\sc ram}.

L'outil de navigation, n\oe{}ud du réseau, comporte en interface utilisateur un indicateur d'utilisation mémoire, ainsi qu'un interupteur servant à enregistrer ou au contraire à désapprendre le contenu des {\sc ig}s rencontrés. La gestion des {\sc ig}s sauvegardés localement est dépendante de la capacité mémoire du matériel de visualisation et de certaines règles de préférence données par l'utilisateur. Néanmoins, en mode connecté, l'utilisateur retrouve bien entendu tous les {\sc ig}s dont il a fait l'achat ainsi que l'ensembles des autres {\sc ig}s disponibles.

La présentation, indexation, publicité, échantillon, résumé, sont fournis par le réseau {\em Internet}, sur des serveurs {\sc http} classiques, sans besoin d'authentification. En revanche, une authentification en {\em https - ssl/tls} ne peut pas servir d'entrée au réseau \net{} car ce dernier requiert une authentification forte et privée (sans serveur appartenant à une autorité), assurant sa sécurité.

Le protocole du réseau \net{} impose une seule connexion active par utilisateur. Ainsi dès l'instant qu'un utilisateur s'authentifie sur un appareil, sa session est automatiquement fermée sur le précédent appareil utilisé pour être reconduite sur le nouveau. Néanmoins, l'architecture distribuée du réseau assure qu'aucun tiers ou qu'aucune organisation ne puisse localiser un individu par l'adresse {\em ip} du n\oe{}ud que ce dernier utilise. L'unicité des sessions, la sauvegarde du contexte et la personnalisation des achats/ventes d'{\sc ig}s font du réseau \net{} un véritable double numérique de chacun. L'appareil connecté qui sera le plus utilisé est assurément le {\em ``smartphone''} d'autant qu'il sert aussi de matériel d'authentification forte pour remplacer un badge ou une clé {\em usb}. L'utilisateur peut ainsi passer d'un appareil à l'autre très rapidement et garder son contexte de navigation ou de production.

Les spécifications, les algorithmes et les programmes qui définissent le protocole complet du réseau \net{} sont obligatoirement sous licence {\em Open-Source} d'une part pour garantir le sécurité des procédures cryptographiques et d'autre part pour vérifier la bonne application des calculs et transactions en {\em cup}. 

\section{Autres approches}

\subsection{La licence globale}
Il y a presque unanimité pour constater que le financement actuel des créateurs/auteurs d'{\sc ig}s n'est pas satisfaisant. Les intermédiaires et multinationales de production ont pratiquement vérouillé le marché et durcis les lois sur les {\em droits d'auteur}, tandis que la population répond à cette menace par un piratage toujours plus important. Jusqu'à présent, les propositions de résolution de ce dilemme allaient dans le sens d'une {\em licence globale}. Selon ce principe, l'\'{E}tat ou-bien un organisme délégué devait rétribuer les artistes avec le fruit d'une taxe particuliaire (audiovisuelle, ligne {\sc adsl},....) ou plus simplement via l'impôt. On peut facilement contester les critères d'allocation décidé par l'Administration, la variabilité d'un pays à l'autre, le manque d'inovation économique, une source induite d'inégalité et d'injustice, la démotivation des artistes car déviée de leur \oe{}uvre pour rechercher des subvensions, la nécessité d'acquérir un statut professionel, et enfin pour les artistes qui ne seraient pas subventionnés, la nécessité de bacler leur travail pour augmenter leur rendement.  
Inversement, exiger des créateurs une production gratuite et non rémunérée n'est pas une solution car elle ne favorise que la création simple, immédiate, impulsive, sans effort ni matière première couteuse et donc sans véritable valeur.\\
Richard Stallman\cite{stallman} à proposé de rémunérer les auteurs sur la base de la racine cubique du nombre de téléchargements, idée qui se rapproche de notre modèle mais ne résoud pas le problème de responsabilité de la mesure du traffic, ni de la distribution des fonds collectés. 

Le modèle économique présenté dans ce papier propose donc une solution sans recourir à une {\em license globale}, ni sortir les artistes du champs d'activité professionnel.
Ce système est bien entendu compatible avec le modèle de la recherche scientifique qui peut continuer à mettre ses productions dans le domaine public et rétribuer ses chercheurs par un investissement public ou privé. Rien n'interdit aux chercheurs de publier sur \net{} avec un prix/revenu nul, montrant explicitement que leur article est du domaine public, et profitant de ce média pour prendre date, pour certifier de l'identité de sa paternité et pour protéger les données de tout effacement.

\subsection{Les intérmédiaires}
Notre modèle économique met directement en relation les artistes/auteurs d'un {\sc ig} avec les consommateurs. L'exigence d'un authentification forte des personnes exclue d'inscrire par ce système toute personne morale (entreprise, association, \ldots). L'auteur d'un {\sc ig} peut partager une partie de la paternité de son \oe{}uvre avec d'autres personnes désignées et selon une pondération non modifiable après la naissance de l'{\sc ig} sur \net{}. La rétribution automatique des auteurs est donc pondérée par ce partage de paternité.
En revanche, il n'y a aucune obligation pour l'auteur principal à déclarer comme co-auteur des personnes qui étaient impliquées comme intermédiaires dans le modèle classique; éditeur, producteur, diffuseur, imprimeur, agent\ldots. Pour des productions demandant un investissement lourd comme des films, il peut être établi un contrat classique entre l'artiste et les partenaires sur la base du nombre d'achat passé directement à l'auteur. Ce denier peut fournir un preuve (document signé électroniquement) de la liste des acheteurs d'un {\sc ig} dont il est propriétaire, mais aucun tier ne peut se procurer directement cette liste sur \net{}. Pour une part de plus en plus grande de la production mondiale d'{\sc ig}, l'auteur peut disposer à moindre coût des instruments de création et de diffusion. Par exemple, un écrivain peut directement produire le document final d'une grande qualité typographique en \LaTeX{}\cite{lamport} ou \TeX{}\cite{knuth}. De même, il est courant de voir des musiciens ou des photographes s'auto-produire. Rien n'interdit à des journalistes ou critiques d'Art de promouvoir sur {\em Internet} tel ou artiste ou tel \oe{}uvre mais la principale différence avec le modèle classique est que la promotion ne se rétribue pas sur la production artistique. Les gains éventuels d'un campagne de publicité sont à gérér à la marge du réseau \net{}, en parallèle des transactions en {\em cup} réalisées exclusivement entre l'auteur et les acheteurs.\\
Notre système est beaucoup plus sain car chaque \oe{}uvre est lancée sur \net{} avec les mêmes chances de succès. Les intermédiaires ne peuvent pas perturber le marché des {\sc ig}s au profit d'une minorité d'individus ou pire diriger, infuencer, acheter les artistes pour contraindre leurs créations à fins mercantiles, et en même temps abrutir la population par une focalisation sélective. On se rend compte que l'inadaptation du modèle classique de commerce aux {\sc ig}s, dont le coût de reproduction est nul, a permis à des parasites non créateur de se nourrir sur le dos de vrais auteurs/artistes. Ces intermédiaires justifaient facilement leur existence avant l'aire {\em Internet} par la main mise sur des réseaux de distribution matériels obligatoires. Malgrès les résistances, la situation change et les intermédiaires vont se cantonner sur le marché des biens matériels, un luxe de la matérialisation des {\sc ig}.\\
Constatons que le réseau \net{} est doublement démocratique par son accès égalitaire pour tout citoyen et par le nouveau modèle économique rétribuant plus justement les artistes, en toute transparence pour la population.
Un avantage suplémentaire du réseau \net{} est qu'en limitant les transactions qu'aux véritables créations, le prix payé par le consommateur final ne peut qu'être inférieur à celui pratiqué dans un système monopolisé et géré par des intermédiaires non humains. Des prix bas ont alors un effet positif sur la consommation de masse. Cet effet est encore démultiplié par notre modèle qui, rappelons le, redistribue équitablement les nouveaux gains entre l'auteur/vendeur et les acheteurs.
\subsection{Le droit d'auteur revu}
La personnalisation requise des acteurs du réseau \net{} définit implicitement un {\em droit d'auteur}, pour la durée de vie de ce dernier. Si l'auteur décède, il est techniquement possible de mettre en place avec le concours des banques un transfert des {\em cup} gagnés vers des ayants droits. Le profit reste somme toute limité dans notre modèle économique dès lors que le revenu prévu maximal a été atteint. Dans ce cas, si l'\oe{}uvre continue à se vendre, les ayants droits ne touchent aucun dividende car les achats sont automatiquement redistribués aux précédents acheteurs. 
Il demeure la question difficile d'évaluer dans chaque {\sc ig} la part d'inspiration prises aux générations précédentes par rapport à la part purement innovante. La qualification de {\em vol} d'{\sc ig} restera surement une appréciation subjective pour la Justice. Rappelons que l'anonymat n'ést pas possible sur \net{}, et que les \oe{}uvres sont horodatés, ce qui simplifie pour partie la tâche des juges.\\
De plus, notre modèle défini une sortie naturelle vers le domaine public quand les acheteurs ne payent quasiment rien, zéro ou un centime de {\em cup} suite à des arrondis de calcul. La fixation d'une date de 70 ans ou 80 ans pour les droits d'auteur n'a plus alors de raison d'être car il n'y a plus à résoudre le problème d'attribution des gains sur des \oe{}uvres anciennes. Enfin, il n'y a plus à forcer une justification artificielle d'un droit du sang qui tranmettrait le génie d'un artiste. Tout se passe comme si les véritables enfants d'un créateur étaient et resteraient la population mondiale des consommateurs. N'est ce pas l'Amour de l'appartenance à notre société le principal moteur de toute création artiste~?
\subsection{La crainte du piratage}
L'histoire de l'Informatique a montré que les systèmes vraiment sécurisés ne le sont pas éternellement et dévoilent souvent quelques failles ou faiblesses. Or l'acceptation du réseau \net{} et du modèle économique associé est entièrement conditionné à une sécurité exemplaire. En particulier, la protection des données privées des citoyens, d'une part la sauvegarde des {\sc ig}s crées ou achetés, ce indéfiniment et d'autre part l'impossibilité pour un tier, même doté d'un status élevé dans l'organisation de l'\'{E}tat, d'obtenir des informations sur une partie de ces {\sc ig}s.  
Tout l'objet de l'organisation nouvellement crée, la {\em ``Fondation$\sqcup$''}\footnote{\url{http://www.cupfoundation.net}} (The {\em ``$\sqcup$\underline{Foundation}''}) est de tester, casser, faire tester la sécurité informatique du réseau avec les techniques les plus avancées de cryptographie.
Un principe immuable reste la diffusion publique des tous les algorithmes et {\em code sources} utilisés par le protocole réseau et les applicatif mis à disposition des utilisateurs. Aucune {\em sécurité du secret} n'est utilisée. Le système est donc protégé par le niveau courant de la recherche en Mathématique et la puissance extrapolées des ordinateurs. Comme le réseau est totalement géré par les n\oe{}uds, sans serveur central, il n'y a pas d'administrateur pouvant avoir plus de droits que le citoyen lambda connectant son {\sc pc} au réseau \net.

\section{Conclusion}
Nous avons présenté une solution possible à l'inadaptation du commerce classique aux biens marchands immatériels. Cette nouvelle Économie est basée sur une transaction de type {\em 1-n}, qui borne le revenu de l'auteur/vendeur, tout en équilibrant dans le temps le coût pour chaque acheteur. Elle implique une création d'une monnaie, le {\em "squarecup"} ou {\em "cup"} noté: $\sqcup$, à deux nombres, ainsi qu'un réseau informatique, \net{} dont le protocole est {\em Open-source}, une architecture distribuée (modèle pair à pair), réseau basé sur une authentification privée, un service de "zéro téléchargement" et une très forte sécurité (chiffrage, signature et horodatage des contenus). Les trois acteurs impliqués dans ce réseaux; les pouvoirs publics, les banques et les citoyens bénéficient par ce système d'une relance de l'activité et d'une amélioration globale de leur revenus. Le réseau \net{} est en phase de prototypage et sa spécification complète devra être gérée par un organisme indépendant, la {\em Fondation$\sqcup$}, qui pourra certifier de la sécurité du système avant son déploiement par {\em Internet}.
\begin{flushright}{\tiny The end of the document}\end{flushright}

\end{document}